\documentclass[10pt,journal,compsoc]{IEEEtran}

\usepackage{epsfig}
\usepackage{graphicx}
\usepackage[tight,footnotesize]{subfigure} %@@ check font size
\usepackage{caption} %@@ define fontsize!
\usepackage{epstopdf}
\usepackage{balance}

\usepackage{color}
\usepackage{xcolor}
\usepackage{colortbl}
\newcommand{\blue}[1]{\textcolor[rgb]{0.00,0.00,0.00}{#1}}

\newcommand{\red}[1]{\textcolor[rgb]{1.00,0.00,0.00}{\textbf{#1}}}

\definecolor{Gray}{gray}{0.9}

\definecolor{linkcol}{rgb}{0,0,0.75}
\definecolor{citecol}{rgb}{0,0,0}
\definecolor{urlcol}{rgb}{0,0,0}

\usepackage{hyperref}

\usepackage{breakurl}
\usepackage{booktabs}
\usepackage{tabularx}

\usepackage{array} 
\newcolumntype{H}{>{\setbox0=\hbox\bgroup}c<{\egroup}@{}}

\newcommand{\ed}[1]{\textsf{\textcolor{red}{[#1]}}}

\usepackage{comment}
\long\def\comment#1{}

\newcommand{\remove}[1]{}

\includecomment{emergencydroppable} % nspring added to demarcate paragraphs that could go if we need them to.  but they're still in.
\excludecomment{droppingdetailforspace} % nspring added to demarcate paragraphs that we'll cut but we may decide to include if we have space (in that case change this definition into \includecomment..)

\providecommand{\ie}{\emph{i.e.,} }
\providecommand{\eg}{\emph{e.g.,} }
\providecommand{\cf}{\emph{cf.,} }

\usepackage{amssymb}
\usepackage{amsmath}
\usepackage{mathrsfs}
\usepackage{mathbbol}
\usepackage{bm}			% bold math even for greek letters

\usepackage{amsthm} % may give problems with sigcomm template

\newcommand{\myitem}[1]{\vspace*{0.04in}\noindent\textbf{#1}}

\usepackage[normalem]{ulem}

\providecommand{\ione}{\emph{(i)} }
\providecommand{\itwo}{\emph{(ii)} }
\providecommand{\ithree}{\emph{(iii)} }
\providecommand{\ifour}{\emph{(iv)} }

\usepackage{cleveref}
\crefformat{footnote}{#2\footnotemark[#1]#3}

\usepackage{verbatim}

\usepackage{listings}

\usepackage{enumitem}
\setlist[description]{leftmargin=*}

\usepackage{cite}
\usepackage{xspace}
\newcommand{\artemis}[0]{\textsf{ARTEMIS}\xspace} % usage: \artemis\ to keep whitespace

\newcommand{\yes}{$\checkmark$}
\newcommand{\no}{$\times$}
\newcommand{\?}{\red{\textbf{?}}}
\newcommand{\NA}{\textit{N/A}}

\def\papertitle{ARTEMIS: Neutralizing BGP Hijacking\\ within a Minute}

\hypersetup{
	pdfauthor = {},
	pdftitle = {\papertitle},
	pdfcreator = {LaTeX with hyperref package},
	pdfproducer = {dvips + ps2pdf}
}

\begin{document}
%\pagenumbering{arabic}
%\setcounter{page}{1}

\title{\papertitle}
%\numberofauthors{1}
%
%\newcommand{\inst}[1]{$^{#1}$}
%\author{
%\hspace{-10pt}
%Pavlos Sermpezis\inst{1}, 
%Vasileios Kotronis\inst{1},
%Petros Gigis\inst{1},
%Xenofontas Dimitropoulos\inst{1,2},
%Jae Hyun Park\inst{3},
%Danilo Cicalese\inst{3,4},
%Alistair King\inst{3},
%Alberto Dainotti\inst{3}
%\\[4pt]
%$^1$FORTH%\\[4pt]
%\hspace{10pt}
%$^2$University of Crete
%\hspace{10pt}
%$^3$CAIDA, UC San Diego
%\hspace{10pt}
%$^4$Telecom ParisTech
%}

\author{Pavlos~Sermpezis, 
Vasileios~Kotronis, 
Petros~Gigis,
Xenofontas~Dimitropoulos,\\
Danilo~Cicalese,
%Jae~Hyun~Park,
Alistair~King,
and~Alberto~Dainotti% <-this % stops a space
\IEEEcompsocitemizethanks{\IEEEcompsocthanksitem 
P. Sermpezis, V. Kotronis, and P. Gigis are with ICS-FORTH, Greece; X.
Dimitropoulos is with ICS-FORTH and University of Crete, Greece; %J.H. Park, 
A. King, and A. Dainotti are with CAIDA, UC San Diego, USA; D. Cicalese is with CAIDA, UC San Diego, CA, USA, and Telecom ParisTech, France.\protect\\
%%%%%%\IEEEcompsocthanksitem% J. Doe and J. Doe are with Anonymous University.%
}% <-this % stops an unwanted space
%\thanks{Manuscript received Month xx, 2017; revised Month xx, 2017.}
}

\IEEEtitleabstractindextext{%
\begin{abstract}
% 150-250 wordsfor ToN -- one paragraph !!
%
%%%%%%%%%%%%%REVISED ABSTRACT BY VK%%%%%%%%%%%%%
BGP prefix hijacking is a critical threat to Internet organizations and users.
Despite the availability of several defense approaches (ranging from RPKI to popular third-party services), none of them solves the problem adequately in practice. In fact, they suffer from%, suffering from
: (i) lack of detection comprehensiveness, allowing sophisticated
attackers to evade detection, (ii) limited accuracy, especially in the
case of third-party detection, (iii) delayed verification and
mitigation of incidents, reaching up to days, and (iv) lack of privacy
and of flexibility in post-hijack counteractions, on the side of network operators.
In this work, %encapsulating feedback from a %comprehensive 
%survey among operators, 
we propose ARTEMIS \blue{(Automatic and Real-Time dEtection and MItigation System)}, a defense approach (a) based on \textit{accurate and fast
detection operated by the AS itself}, leveraging the pervasiveness of publicly available BGP monitoring services and their recent shift towards real-time streaming, thus (b) enabling \textit{flexible and fast mitigation} of hijacking events. 
Compared to previous work, our approach
combines characteristics desirable to network operators such as
comprehensiveness, accuracy, speed, privacy, and flexibility. Finally, we show through real-world experiments that, with the ARTEMIS approach, prefix hijacking can be neutralized within a minute.

\end{abstract}

%\begin{IEEEkeywords}
%Computer Society, IEEE, IEEEtran, journal, \LaTeX, paper, template.
%\end{IEEEkeywords}
}

\maketitle

\IEEEraisesectionheading{\section{Introduction}\label{sec:introduction}}
\IEEEPARstart{A}{utonomous} Systems (ASes) use the Border Gateway Protocol
(BGP)~\cite{BGPv4} to advertise their IP prefixes and establish
inter-domain routes in the Internet. BGP is a distributed
protocol, lacking authentication of routes. As a
result, an AS is able to advertise illegitimate routes for
IP prefixes it does not own. These illegitimate advertisements propagate and ``pollute'' many
ASes, or even the entire Internet, 
affecting service availability, integrity, and confidentiality of
communications.
This phenomenon, called \textit{BGP prefix hijacking} can be caused by
router misconfiguration~\cite{hijack-YouTube,hijack-ChinaTelecom} or
malicious
attacks~\cite{hijack-BitCoins,Ramachandran-BGP-spammers-CCR-2006,
Vervier-mind-blocks-NDSS-2015}. 
Events with significant impact are
frequently
observed~\cite{russian-hijack-2017, iranian-hijack-2017, backconnect-hijack-2016, Vervier-mind-blocks-NDSS-2015}, highlighting -- despite
the severity of such Internet infrastructural vulnerability --
the ineffectiveness of existing countermeasures. 
%@ better to state this here before next paragraph, so the reader
% reads the next paragraph not misunderstanding that the practical
% solutions have solved the problem

Currently, networks rely on \textit{practical reactive mechanisms} to
try to
defend against prefix hijacking, since proposed \textit{proactive}
mechanisms~\cite{Kent-secure-BGP-JSAC-2000,Subramanian-listen-whisper-NSDI-2004,bgpsec-specification-2015,rpki-rfc,Karlin-PGBGP-ICNP-2006}
(\eg RPKI)
are fully efficient only when globally deployed, and operators are
reluctant
to deploy them due to associated technical and financial costs~\cite{lychev-2013-sigcomm,cooper-2013-hotnets,matsumoto-2017-privsec}.
Defending against hijacking reactively consists of two steps:
\textit{detection} and \textit{mitigation}. Detection is mainly
provided by third-party services, \eg \cite{commercial-bgpmon}, that
notify networks about suspicious events involving their prefixes,
based on routing information (such as traceroutes~\cite{Zheng-lightweight-hijacks-2007} or BGP updates~\cite{commercial-bgpmon}).
The affected networks then proceed to mitigate the event, \eg by announcing more specific
prefixes, or contacting other ASes to filter announcements. 

However, due to a mix of technological and practical deployability issues,
current reactive approaches are largely inadequate.
In this paper, we address these issues by proposing 
\artemis (Automatic and Real-Time dEtection and MItigation
System), a \textit{self-operated} and \textit{unified} detection and
mitigation approach  based on \textit{control-plane monitoring}.
Specifically, the state of the art suffers from 4 main problems:
%, all of which we address in this paper:

\begin{itemize}[leftmargin=*]
%\item \textbf{Comprehensiveness}.
\item \textbf{Evasion}.
%\item \textbf{Attack Resistance}.
None of the detection approaches in literature is 
capable of detecting all attack configurations 
(nor can they be easily
combined), thus allowing sophisticated
attackers to evade them. We propose a modular taxonomy
describing all variations of attack scenarios
%(\S~\ref{sec:hijack-types-characteristics})
and we use it to
carefully analyze detection comprehensiveness of related work.
%(\S~\ref{sec:related}).
%and to demonstrate that 
%On the contrary, 
%The solution we propose -- \artemis (\textit{Automatic and Real-Time dEtection and MItigation System}) -- 
\textit{\artemis
significantly overcomes limitations of the state of the art by
covering all 
attack configurations in the context of a common threat model}.
% (\S~\ref{sec:detection}).

\item \textbf{Accuracy}. 
%Real-world implementations currently used
%for detection are based on third-party services. 
Legitimate changes in the routing policies of a network (\eg 
announcing a sub-prefix for traffic engineering or establishing
a new peering connection), could be considered suspicious events
by the majority of third-party detection
systems~\cite{Chi-cyclops-CCR-2008,Lad-Phas-Usenix-2006,Zheng-lightweight-hijacks-2007,Shi-Argus-IMC-2012,Hu-accurate-hijacks-SP-2007}.
To avoid this, operators would need to timely inform third parties
about every routing decision they make and share private information.
%(\eg ISPs usually do not disclose their peering policies). 
On the other hand, adopting a less strict policy to compensate for the
lack of updated information and reduce false positives (FP), incurs
the danger of neglecting real hijacking events (false negatives --
FN). \textit{We designed \artemis detection
%(\S~\ref{sec:detection})
to be run directly by the network operator
without relying on a third party, thus leveraging fully and
constantly (and potentially automatically) updated information that
enables 0\% FP and FN for most of the attack scenarios and a
configurable FP--FN trade-off otherwise}. %for the remaining ones.
%potentially malicious events with negligible impact.

\item \textbf{Speed}.  A side effect of the inaccuracy of third-party
approaches 
%(\S~\ref{sec:motivation}) 
is the need for manual verification of alerts, which
inevitably causes slow mitigation of malicious events (\eg hours or days). 
Few minutes of diverted traffic can cause large financial losses due to 
service unavailability or security breaches.
%lack of full automation However, this widely followed approach
%typically in- volves significant delay until the mitigation of a
%hijacking event, reaching several hours or even days. This delay is
%mainly due to the manual actions needed for the (i) verification of
%alerts, and (ii) mitigation of the events.  Manual verification of
%alerts is a side-effect of the fact that third-party detection can be
%inaccurate.
On the contrary, \textit{\artemis is a fully automated solution integrating detection
and mitigation, allowing an AS to quickly neutralize attacks}.
%The \artemis approach is immediately
%deployable \emph{today}: we build a prototype system implementing our approach, and
We
 conduct real hijacking experiments in the Internet
demonstrating that \artemis can detect attacks \textit{within
seconds} and \textit{neutralize them within a
minute, i.e., orders of magnitude faster than
current practices}. 
%(\S~\ref{sec:experiments}).

\item \textbf{Privacy and Flexibility}.
One of the issues that impedes the adoption of third-party detection
is privacy, \eg ISPs usually do not disclose their peering policies.
%(\S\ref{sec:motivation}). 
Similarly, operators are sometimes reluctant to adopt mitigation services 
requiring other organizations to announce their prefixes or 
tunnel their traffic.
%(\S\ref{sec:motivation}). 
\textit{\artemis offers full privacy for detection and the option to achieve
self-operated mitigation}.
Another factor affecting willingness to 
externalize mitigation is cost.
%(\S\ref{sec:motivation}). 
Trade-offs between cost, privacy, and
risk may be evaluated differently by the same organization for
distinct prefixes they own.
\textit{Leveraging the availability of local private information and its fully
automated approach, 
\artemis offers the flexibility to customize mitigation (\eg
self-operated or third-party-assisted) per prefix and per attack class}.
%We analyze such real-world deployment issues and impediments also 
%through the help of a ... survey among operators..

\end{itemize}

%The novelty of \artemis relies on two key observations
\blue{The \artemis approach relies on two key observations}: \ione
today's public BGP monitoring infrastructure (such as RouteViews~\cite{routeviews} and
RIPE RIS~\cite{ripe-ris-real-time}) is much more advanced than when previous solutions for BGP
hijacking detection were proposed, making it a valuable resource for
comprehensive, live monitoring of the Internet control plane that is available to
anybody; \itwo shifting from a third-party perspective to a
self-operated approach enables us to effectively address the
long-standing and persistent issues undermining the state of the art in
BGP hijacking defense approaches.

In this work, we first define our threat model and propose a novel attack taxonomy used throughout the paper (\S~\ref{sec:hijack-types-characteristics}).
%We evaluate our design decisions with respect to detection and mitigation of BGP hijacking through simulations and analysis of real-world Internet control-plane measurements (\S~\ref{sec:data},\S~\ref{sec:understanding-hijacks},\S~\ref{sec:detection},\S~\ref{sec:mitigation-techniques}).
\blue{We investigate the visibility (from the public monitoring infrastructure \S~\ref{sec:data}) and impact of different hijacking types in \S~\ref{sec:understanding-hijacks}, and then design the \artemis detection (\S~\ref{sec:detection}) and mitigation (\S~\ref{sec:mitigation-techniques}) approach. We evaluate our design decisions through simulations and analysis of real-world Internet control-plane measurements (\S~\ref{sec:data},\S~\ref{sec:understanding-hijacks},\S~\ref{sec:detection},\S~\ref{sec:mitigation-techniques}).}
%\blue{We evaluate our design decisions with respect to detection (\S~\ref{sec:detection}) and mitigation (\S~\ref{sec:mitigation-techniques}) of BGP hijacking through simulations and analysis of real-world Internet control-plane measurements (\S~\ref{sec:data},\S~\ref{sec:understanding-hijacks},\S~\ref{sec:detection},\S~\ref{sec:mitigation-techniques}).}
Furthermore, the \artemis approach is immediately deployable
\emph{today}: we build a prototype system implementing our approach,
and we show its effectiveness through experiments on the real
Internet (\S~\ref{sec:experiments}).
%Finally, we provide extensive background pertaining to our solution through a deep analysis of the state of the art, both in terms of practical experience -- by conducting a survey among operators and referring to documented events -- (\S~\ref{sec:motivation}) and related literature (\S~\ref{sec:related}).
\blue{Finally, we provide an extensive background on the state of the art, both in terms of practical experience (by conducting a survey among operators and referring to reported events; \S~\ref{sec:motivation}) and related literature (\S~\ref{sec:related}).}
%This work represents our first step and a reference towards developing a fully operational open-source tool that could be easily deployed by operators worldwide, with the potential to dramatically reduce the impact of BGP prefix hijacking on the Internet.

%\section{Background and Problem Scope}
%\label{sec:preliminaries}
%\input{Preliminaries}

\section{Threat Model and Attack Taxonomy}
%\subsection{Threat Model and Attack Taxonomy}
\label{sec:hijack-types-characteristics}
%\subsection{BGP Hijacking (Threat) Model}
We consider a common and general hijacking threat model (\eg similarly to ~\cite{heap-jsac2016}), where a hijacker controls a single AS and its edge routers, and has full control of the control plane and data plane within its own AS. The hijacker can arbitrarily manipulate the BGP messages that it sends to its neighboring ASes (control plane) and the traffic that crosses its network (data plane), but has otherwise no control over BGP messages and traffic exchanged between two other ASes.

In this threat model, there are three dimensions that
characterize how a hijacking attack can be carried out: \ione the
affected prefix, \itwo the manipulation of the \texttt{AS-PATH} in the BGP
messages, and \ithree how the (hijacked) data-plane traffic is
treated. Any attack can be represented by a ``point'' in this
three-dimensional ``space''.
\blue{Table~\ref{table:taxonomy-related-work} presents all possible
attack combinations (three leftmost columns); ``*'' denote wildcarded fields.}

\begin{table*}
\centering
\caption{Comparison of BGP prefix hijacking detection systems/services w.r.t. ability to detect different classes of attacks.}
\vspace{-3mm}
\label{table:taxonomy-related-work}
\begin{tabular}{|ccc|c|c|c|c|c|c|c|c|}
\hline
\rowcolor{Gray}
%\multicolumn{3}{|c|}{Class of}			& \multicolumn{8}{c|}{Detection System/Service} \\
%\rowcolor{Gray}
\multicolumn{3}{|c|}{Class of Hijacking Attack}	& \multicolumn{3}{c|}{Control-plane System/Service} & \multicolumn{2}{c|}{Data-plane System/Service} & \multicolumn{3}{c|}{Hybrid System/Service}\\
\hline
\rowcolor{Gray}
{Affected}	& {AS-PATH}	& {Data}	& {\artemis}		&{Cyclops}	& {PHAS} & {iSpy}	& {Zheng \textit{et al.}}	& {HEAP}	& {Argus}	& {Hu \textit{et al.}}	\\
\rowcolor{Gray}
{prefix}		& {(Type)}			& {plane}& {}&{(2008)~\cite{Chi-cyclops-CCR-2008}}	& {(2006)~\cite{Lad-Phas-Usenix-2006}} & {(2008)~\cite{Zhang-Ispy-CCR-2008}}	& {(2007)~\cite{Zheng-lightweight-hijacks-2007}}	& {(2016)~\cite{heap-jsac2016}}	& {(2012)~\cite{Shi-Argus-IMC-2012}}	& {(2007)~\cite{Hu-accurate-hijacks-SP-2007}}	\\
\hline
{Sub}&{U}&{*} 					&{\yes}&{\no}&{\no}&{\no}&{\no}&{\no}&{\no}&{\no}\\
\hline
{Sub}&{0/1}&{BH} 				&{\yes}&{\no}&{\yes}&{\no}&{\no}&{\yes}&{\yes}&{\yes}\\
\hline
{Sub}&{0/1}&{IM} 				&{\yes}&{\no}&{\yes}&{\no}&{\no}&{\yes}&{\no}&{\yes}\\
\hline
{Sub}&{0/1}&{MM} 				&{\yes}&{\no}&{\yes}&{\no}&{\no}&{\no}&{\no}&{\no}\\
\hline
{Sub}&{$\geq2$}&{BH}		    &{\yes}&{\no}&{\no}&{\no}&{\no}&{\yes}&{\yes}&{\yes}\\
\hline
{Sub}&{$\geq2$}&{IM}		    &{\yes}&{\no}&{\no}&{\no}&{\no}&{\yes}&{\no}&{\yes}\\
\hline
{Sub}&{$\geq2$}&{MM}		    &{\yes}&{\no}&{\no}&{\no}&{\no}&{\no}&{\no}&{\no}\\
\hline
{Exact}&{0/1}&{BH}				&{\yes}&{\yes}&{\yes}&{\yes}&{\no}&{\no}&{\yes}&{\yes}\\
\hline
{Exact}&{0/1}&{IM}				&{\yes}&{\yes}&{\yes}&{\no}&{\yes}&{\no}&{\no}&{\yes}\\
\hline
{Exact}&{0/1}&{MM}				&{\yes}&{\yes}&{\yes}&{\no}&{\yes}&{\no}&{\no}&{\no}\\
\hline
{Exact}&{$\geq2$}&{BH}		    &{\yes}&{\no}&{\no}&{\yes}&{\no}&{\no}&{\yes}&{\yes}\\
\hline
{Exact}&{$\geq2$}&{IM}		    &{\yes}&{\no}&{\no}&{\no}&{\yes}&{\no}&{\no}&{\yes}\\
\hline
{Exact}&{$\geq2$}&{MM}		    &{\yes}&{\no}&{\no}&{\no}&{\yes}&{\no}&{\no}&{\no}\\
\hline
\end{tabular}
\end{table*}

In the following, we provide a taxonomy of hijacking attacks,
based on these 3 properties.
%Fig.~\ref{fig:taxonomy-hijacks} visualizes the three-dimensional space that defines a hijacking attack.
%
%\begin{figure}
%\centering
%\includegraphics[width = 0.8\linewidth, trim=3.8cm 5cm 3.8cm 2.5cm,clip]{./figures/3d-hijack-taxonomy}
%\caption{Taxonomy of BGP prefix hijacking attacks/incidents: a hijack can be any point in the three dimensional space of the \{Prefix, AS-PATH, Data-plane\} properties.}
%\label{fig:taxonomy-hijacks}
%\end{figure}
%
%\subsection{Taxonomy of BGP Hijacks}
For the sake of demonstration, we assume that \textit{AS1} owns and
legitimately announces the prefix \textit{10.0.0.0/23}, and
\textit{AS2} is the hijacking AS. We denote a BGP message with two
fields: its \texttt{AS-PATH} and announced prefix. For example,  
\textit{\{ASx, ASy, AS1 -- 10.0.0.0/23\}}
is a BGP announcement for prefix \textit{10.0.0.0/23}, with \texttt{AS-PATH}
\textit{\{ASx, ASy, AS1\}}, originated by the legitimate AS
(\textit{AS1}).

\vspace{-4pt}
\subsection{Classification by Announced AS-Path}
\label{sec:taxonomy-as-path}
%\vspace{-6pt}
\vspace{-4pt}
\myitem{Origin-AS (or Type-0) hijacking:} The hijacker \textit{AS2}
announces -- as its own -- a prefix that it is not authorized to originate,
\eg \textit{\{AS2 -- 10.0.0.0/23\}}. This is the most
commonly observed hijack type, and might occur either due to an attack or
a misconfiguration. 

\myitem{Type-N hijacking ($N\geq1$):} The hijacker \textit{AS2} deliberately
announces an illegitimate path for a prefix it does not own. %The path contains the ASN of the hijacker as the last hop, 
\blue{The announced path contains the ASN of the victim (first AS in the path) and hijacker (last AS in the path),} \eg \textit{\{AS2,
ASx, ASy, AS1 -- 10.0.0.0/23\}}, while the sequence of ASes in the path
is not a valid route, \eg AS2 is not an actual neighbor of ASx.
In our taxonomy, the position of the \emph{rightmost fake link} in the forged
announcement determines the \textit{type}. E.g., 
\textit{\{AS2, AS1 -- 10.0.0.0/23\}} is a \textit{Type-1}
hijacking, \textit{\{AS2, ASy, AS1 -- 10.0.0.0/23\}} is a
\textit{Type-2} hijacking, etc.

\myitem{Type-U:} The hijacker leaves the legitimate \texttt{AS-PATH}
unaltered (but may alter the announced prefix~\cite{defcon16-attack})
\footnote{\blue{If the announced prefix is also left unaltered (\ie no path or prefix manipulation; see \S~\ref{sec:taxonomy-affected-prefix}), then the event is not a hijack (no misuse of BGP) but a traffic manipulation attempt, out of the scope of this paper.}}
%\footnote{\blue{Note that a Type-U attack with unaltered prefix(es), i.e., an exact-prefix hijack (\S~\ref{sec:taxonomy-affected-prefix}), is essentially a traffic manipulation attack (\S~\ref{sec:taxonomy-data-plane}), and not a hijack that is visible on the control plane.}}
. 

\vspace{-4pt}
\subsection{Classification by Affected Prefix}
\label{sec:taxonomy-affected-prefix}
%\vspace{-6pt}
\vspace{-4pt}

\myitem{Exact prefix hijacking:} The hijacker announces a path for
exactly the same prefix announced by the legitimate AS. Since
shortest AS-paths are typically preferred, only a part of
the Internet that is close to the hijacker (\eg in terms of
AS hops) switches to routes towards the hijacker. The examples
presented above (\S~\ref{sec:taxonomy-as-path}) are exact prefix hijacks.

\myitem{Sub-prefix hijacking:} The hijacker \textit{AS2} announces a more specific
prefix, \ie a sub-prefix of the prefix of the legitimate AS.
For example, \textit{AS2} announces a path \textit{\{AS2 -- 10.0.0.0/24\}} or \textit{\{AS2, ASx, ASy, AS1 -- 10.0.0.0/24\}}.
%(the path announced by the hijacker might match its legitimate
%path for the prefix 10.0.0.0/23~\cite{defcon16-attack}). 
Since in BGP more specific prefixes are preferred, \textit{the entire
Internet} routes traffic towards the hijacker to reach the announced
sub-prefix.
% contributes around \red{10\%} of all stable hijackings in the Internet~\cite{Shi-Argus-IMC-2012}. 

\myitem{Squatting:} The hijacker AS announces a prefix owned but not (currently) announced by the owner AS~\cite{nanog-squatting}. 

\vspace{-4pt}
\subsection{Classification by Data-Plane Traffic Manipulation}
\label{sec:taxonomy-data-plane}
%\vspace{-6pt}
\vspace{-4pt}

The effect of a hijack is to redirect traffic for the affected prefix
to/through the network of the hijacker AS. This attracted traffic can
be \ione dropped (\textit{blackholing, BH}), \itwo manipulated or
eavesdropped and then sent on to the victim \textit{AS1}
(\textit{man-in-the-middle, MM}), or \ithree used in an impersonation
of the victim's service by responding to the senders
(\textit{imposture, IM}). While BH attacks might be easily noticed in
the data plane (since a service is interrupted), MM or IM attacks can
be invisible to the victim AS or the other involved ASes.

\vspace{-4pt}
\subsection{\blue{Example Hijack Scenarios \& Motivations}}
\label{sec:taxonomy-scenarios}
\vspace{-4pt}

%\blue{Here, we illustrate examples of different hijack scenarios with different hijacker motivations.}
\blue{The following examples illustrate different hijack scenarios, their underlying motivation, and how they are classified according to the presented taxonomy.}

%\myitem{\blue{Human Error:}} \blue{In this scenario, the hijack is not deliberate, but the result of a routing misconfiguration. A prominent example is the leakage of a full BGP table from Google to Verizon in 2017~\cite{google-japan-leak}, serving as an accidental large-scale type-0 sub-prefix hijack, with blackholing on the data plane.}
\myitem{\blue{Human Error.}} \blue{The hijack is the result of a routing misconfiguration; \eg the leakage of a full BGP table from China Telecom~\cite{hijack-ChinaTelecom}, led to an accidental large-scale Type-0 exact-prefix hijack, with blackholing on the data plane.}

%\myitem{\blue{High Impact, High Visibility Attack:}} \blue{In this scenario, the hijack is intentional and highly visible with widespread impact. A classic example is the incident involving YouTube and Pakistan Telecom~\cite{ripe-pakistan} where the latter network, aiming to block the YouTube website, engaged in a type-0 sub-prefix hijack (with blackholing on the data plane) affecting YouTube's services for 2 hours worldwide.}
\myitem{\blue{High Impact Attack.}} \blue{The hijack is intentional, with widespread impact; \eg Pakistan Telecom engaged in a Type-0 sub-prefix hijack, blackholing YouTube's services for approximately 2 hours worldwide~\cite{ripe-pakistan}.}%. A classic example is the incident involving YouTube and Pakistan Telecom~\cite{ripe-pakistan} where the latter network engaged in a Type-0 sub-prefix hijack, blackholing YouTube's services for approximately 2 hours worldwide.}

%\myitem{\blue{Variable/Targeted Impact, Stealthy Attack:}} \blue{In this scenario, the hijacking AS launches a very targeted attack, attempting e.g., to steal data as a man-in-the-middle on the data plane, while remaining under the radar with a Type-N or Type-U (\S~\ref{sec:taxonomy-as-path}) attack on the control plane. An example is the Russian hijack of traffic destined to Visa and Mastercard in 2017~\cite{russian-hijack-2017}.}
\myitem{\blue{Targeted, Stealthy Attack.}} \blue{The hijacking AS launches a very targeted attack, attempting to intercept traffic (man-in-the-middle), while remaining under the radar on the control plane (Type-N or Type-U attack); \eg a Russian network hijacked traffic destined to Visa and Mastercard in 2017~\cite{russian-hijack-2017}.}

\myitem{\blue{The ``Best'' Attack.}} \blue{Motivations behind hijacks
differ; there is no one ``best'' attack type that is always preferred. For example, an attacker may resort to a Type-N ($N>0$) hijack to \ione evade simple detection systems currently used by operators or bypass RPKI ROV, or \itwo delay manual investigation and recovery from the malicious event; in contrast to origin AS validation, inferring that a link in an AS-path is fake is a hard challenge. Moreover, while a sub-prefix Type-U hijack can be very effective, it might be neither possible nor ideal in some cases; \eg the upstream providers of a hijacker might be configured to not accept routes for prefixes not owned by their customers.}

\section{Datasets and Tools}
\label{sec:data}

%\myitem{\artemis real-time monitoring can make use of existing public
%data sources and therefore it is immediately deployable.}

%Our methodology uses {\em live} BGP feeds from routers distributed
%globally and located in different ASes. In this paper, we evaluate  
%\artemis using two publicly available services
%that offer control-plane monitoring from
%many vantage points all over the world, namely the streaming
%service of RIPE RIS~\cite{ripe-ris-real-time} 
%and BGPmon~\cite{bgpmon} (XXX and XXX monitors respectively).
%In addition, in order to understand the effect of adding more data
%sources, we perform additional simulations 
%including BGP feeds from RouteViews monitors (XXX monitors)
%\cite{routeviews}. Several RouteViews monitors have started
%experimentally deploying live BMP~\cite{bmp}
%feeds~\cite{bgp-hackathon-2016} accessible through CAIDA's
%BGPStream~\cite{bgpstream}, therefore it is foreseeable 
%that in the near future more live BGP feeds will be publicly
%available.

%In this section, we describe the control-plane monitoring services
%that we consider (\S~\ref{sec:control-plane-monitoring}), as well
%as the simulation methodology we employ in our study (\S~\ref{sec:simulations}).

\subsection{Control-Plane Monitoring}
\label{sec:control-plane-monitoring}

We study BGP prefix hijacking and evaluate \artemis using publicly
available services that offer control-plane monitoring from multiple
\textit{monitors} worldwide. We define as monitors the ASes that peer
through their BGP routers with the infrastructure of the monitoring
services, and provide BGP feeds (\ie BGP updates and RIBs). 
We consider the following monitoring services and tools.
  
\myitem{BGPmon~\cite{bgpmon}} (from Colorado State
University\footnote{BGPmon is also the name of a commercial 
network monitoring service. Throughout this paper, BGPmon
refers to the (free) service provided by Colorado State
University, unless stated otherwise.}) provides
\textit{live} BGP feeds from several BGP routers of (a) the RouteViews~\cite{routeviews} sites, and (b) a few dozens of peers worldwide.
 
\myitem{RIPE RIS~\cite{ripe-ris}.} RIPE's Routing Information
System (RIS) has $21$ route collectors (RCs) distributed
worldwide, collecting BGP updates from around $300$
peering ASes. Currently, $4$ RCs provide \textit{live} BGP
feeds (from approx. $60$ monitors)~\cite{ripe-ris-real-time}, while
data from all RCs can be accessed (with a delay of a few minutes)
through RIPEstat~\cite{ripestat} or the tools of CAIDA's
BGPStream~\cite{bgpstream-paper, bgpstream-website} framework. However, RIPE RIS is in the process of upgrading all its RCs towards providing real-time BGP feeds~\cite{ripe-ris-goes-streaming}.
 
\myitem{RouteViews~\cite{routeviews}} provides control-plane
information collected from $19$ RCs that are connected to nearly $200$
ASes worldwide. A subset of the RouteViews RCs provide \textit{live}
BGP feeds (through BGPmon), while all data can be accessed with a
delay of approx. 20min (using tools from BGPStream).  Several
RouteViews monitors have started experimentally deploying live
BMP~\cite{scudder2016bgp} feeds~\cite{bgpstream-v2-beta}
%~\cite{bgp-hackathon-2016} 
accessible through BGPStream; it is thus foreseeable that in the near future more live BGP feeds will be publicly available.
 
%\myitem{BGPStream~\cite{bgpstream}} is an open-source framework for live (and historical) BGP data analysis. It enables users to quickly inspect raw BGP data through an API. BGPStream provides access to RouteViews and RIPE RIS data archives, with a delay of $5$min and $20$min, respectively.

Our \artemis prototype employs \textit{live} BGP feeds such as the
BGPmon and RIPE RIS \textit{streaming services}. However, to
understand the effect of adding more data sources, we perform
additional simulations and real data analysis including BGP feeds from
\textit{all the monitors} of RIPE RIS and RouteViews services, which
we access through the API of BGPStream\footnote{In our simulations we
consider only the full-feed monitors~\cite{bgpstream-paper} of RIPE
RIS and RouteViews that are more reliable: we include only full-feed
monitors that consistently provided data during March 2017.}. A
summary of the monitoring services \blue{that we use in this paper} is given in Table~\ref{table:monitoring-services}.

\begin{table}
\begin{minipage}{\linewidth}
\centering
\caption{Control-plane monitoring services}
\label{table:monitoring-services}
%\begin{small}
\begin{tabular}{|l|l|cc|}
\hline
\multicolumn{2}{|c|}{}												&{\#monitors}	&{delay}\\
\hline
{Stream}				&{BGPmon~\cite{bgpmon}}							&{~~8}			&{$<1s$}\\				
{services}			&{RIPE RIS (stream)~\cite{ripe-ris-real-time}}	&{~57}			&{$<1s$}\\
\rowcolor{Gray}
{}					&{Total (unique)}								&{~65}			&{}\\
\hline
\multicolumn{4}{c}{}\\
\hline
{All services}		&{RouteViews~\cite{routeviews}}					&{128}			&{$\sim 20min$}\\
{(BGPStream)}		&{RIPE RIS~\cite{ripe-ris}}						&{120}			&{$\sim 5min$}\\
\rowcolor{Gray}
{}					&{Total (unique)}								&{218}			&{}\\
\hline
\end{tabular}
%\end{small}
\end{minipage}
\end{table}

\subsection{Simulation Methodology}
\label{sec:simulations}
In this paper, through extensive simulations, we evaluate the impact
of different types of hijacks, the performance of the monitoring
services, and the efficiency of various
mitigation methods. To simulate the Internet routing system,
we use a largely adopted methodology~\cite{Let-the-market-BGP-sigcomm-2011,how-secure-goldberg-ComNet-2014,Jumpstarting-BGP-sigcomm-2016,RPKI-deployment-2016}:
we build the Internet topology graph from a large experimentally
collected dataset~\cite{AS-relationships-dataset}, use classic
frameworks for inferring routing policies on existing
links~\cite{stable-internet-routing-TON-2001}, and simulate BGP
message exchanges between connected ASes. 
%We then use our BGP simulator to simulate BGP message exchanges (announcements, withdrawals, updates) between connected ASes.

%\subsubsection
\myitem{\textbf{Building the Internet Topology Graph.}}
We use CAIDA's AS-relationship dataset~\cite{AS-relationships-dataset}, which is collected based on the methodology of~\cite{AS-relationships-IMC-2013} and enriched with many extra peering (p2p) links~\cite{multilateral-peering-conext-2013}. The dataset contains a list of AS pairs with a peering link, which is annotated based on their relationship as \textit{c2p} (\textit{customer to provider}) or \textit{p2p} (\textit{peer to peer}). %In our simulations we use the dataset collected in Nov. 2016. We further observed that due to the relative stability of the AS-level topology, our insights remain consistent against newer datasets too.
In this topology, we represent the monitors of \S~\ref{sec:control-plane-monitoring} as AS nodes using their associated ASNs.

%\subsubsection
\myitem{\textbf{Simulating Routing Policies.}}
When an AS learns a new route for a prefix (or, announces a new prefix), it updates its routing table and, if required, sends BGP updates to its neighbors. The update and export processes are determined by its routing policies. In our simulator, and similarly to previous works~\cite{Let-the-market-BGP-sigcomm-2011,how-secure-goldberg-ComNet-2014,Jumpstarting-BGP-sigcomm-2016,RPKI-deployment-2016}, we select the routing policies based on the classic Gao-Rexford conditions that guarantee global BGP convergence and stability~\cite{stable-internet-routing-TON-2001}.% (omitted here for brevity).

\section{Impact and Visibility}
\label{sec:understanding-hijacks}
In this section, through simulation,
we first study the potential
impact of different hijacking types on the control plane, \ie
their ability to pollute the routing tables of other ASes. We then
evaluate the potential of BGP monitoring services (\eg
RouteViews) to observe
these events. Our simulations suggest that the \textit{current BGP
monitoring infrastructure is able to observe all the events with
significant impact}. These results 
%for the impact of hijacking events
help us design our detection approach (\S~\ref{sec:detection}) and
inform our flexible mitigation approach (\S~\ref{sec:mitigation-techniques}).

\subsection{Impact of Hijacks on the Control Plane}\label{sec:impact-of-hijacks}

An AS receiving routes from two different neighboring ASes for the
same prefix, selects one of them to route its traffic. This path
selection is based on peering policies, local preferences, and
the \texttt{AS-PATH} lengths of the received routes. As a result, the impact of an \textit{exact prefix hijacking} event on the control plane depends on such routing selections. 
To understand how the impact of these events can vary, we perform
simulations on the AS-level topology of the Internet (see
\S~\ref{sec:data}). For each scenario, we simulate $1000$ runs
with varying \{legitimate-AS, hijacker-AS\} pairs\footnote{$1000$
simulation runs provide significant statistical accuracy (\ie small confidence intervals for mean/median values) in all our scenarios.}. We refer to an AS as
\textit{polluted} if it selects a path that contains the ASN of the
hijacker. To quantify the impact of a hijack, we calculate the
\textit{fraction of ASes polluted by the event}, excluding those ASes
that were already polluted before the hijack (\eg customers of
the hijacker AS that always route traffic through it). 
We limit the analysis in this section to exact prefix hijacking, since
\textit{sub-prefix hijacking} pollutes the entire Internet (\S~\ref{sec:taxonomy-affected-prefix}).
%@sorry but the 1st sentence is so unclear that is dangerous. I don't
%think we need to say anything. If we really want to say the second
%one, then it belongs to the paragraph of "motivation" (now 8.1) that
%mentions RPKI
%\blues{
%Due to the limited deployment of RPKI (\eg only 7\% of prefixes with ROA certificates; see Section~\ref{sec:motivation}), we consider hijacking events affecting prefixes not covered by ROAs. Note that RPKI would reduce the impact only for Type-0 and/or sub-prefix hijacking events.
%}

%\red{We limit this analysis to \textit{exact prefix hijacking}, since sub-prefix hijacking pollutes the entire Internet in the absence of proactive defense mechanisms}.

%\ed{I think that this is wrong here, we use reactive mechanisms to deal with it ourselves! --vk}
%mitigation. % mechanisms.
% (such as RPKI/ROA for the prefix, or filtering by the hijacker's
% providers).usingusing

\begin{figure}
\centering
\subfigure[]{\includegraphics[width = 0.51\linewidth]{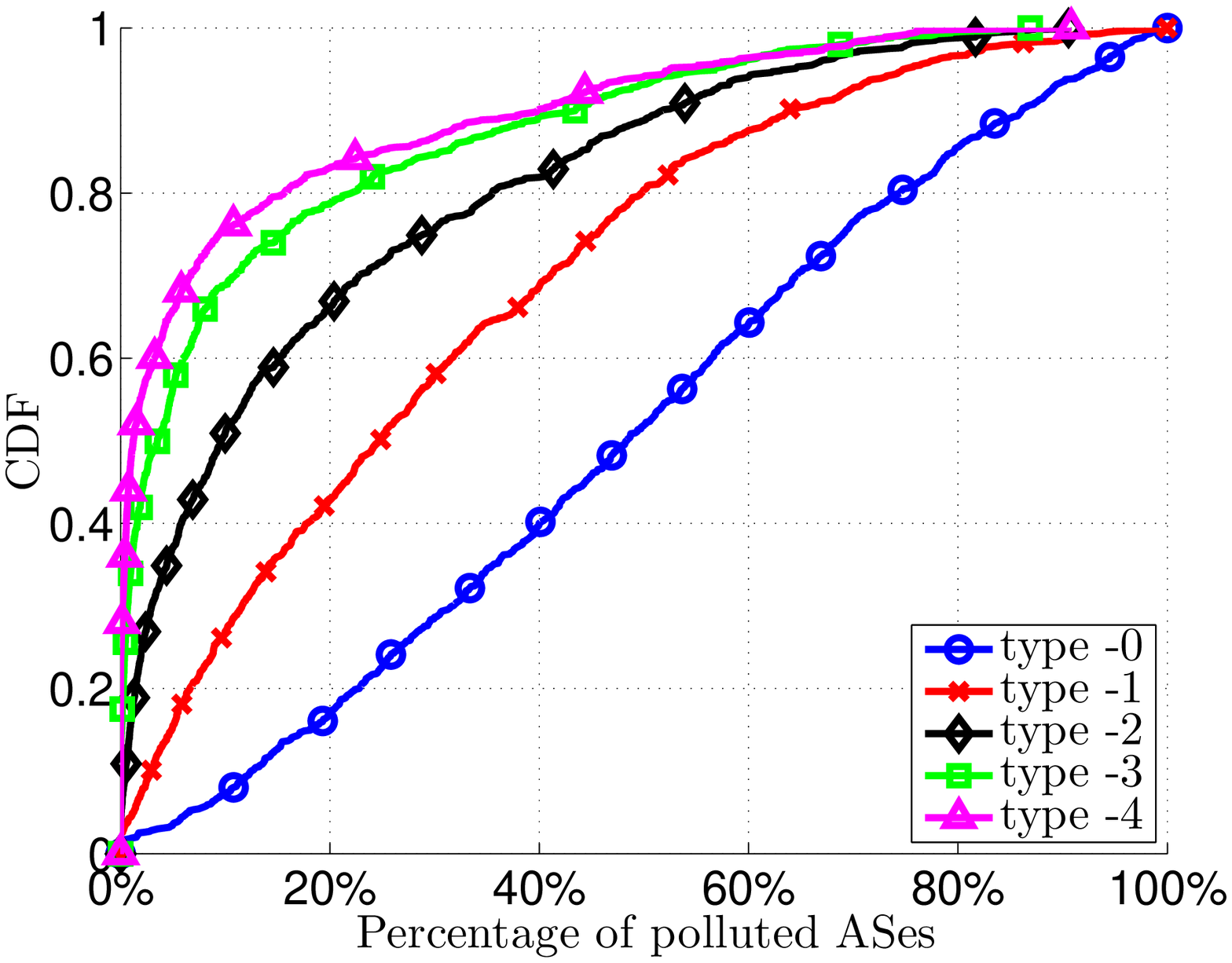}\label{fig:hijack-type-impact}}
\subfigure[]{\includegraphics[width = 0.43\linewidth]{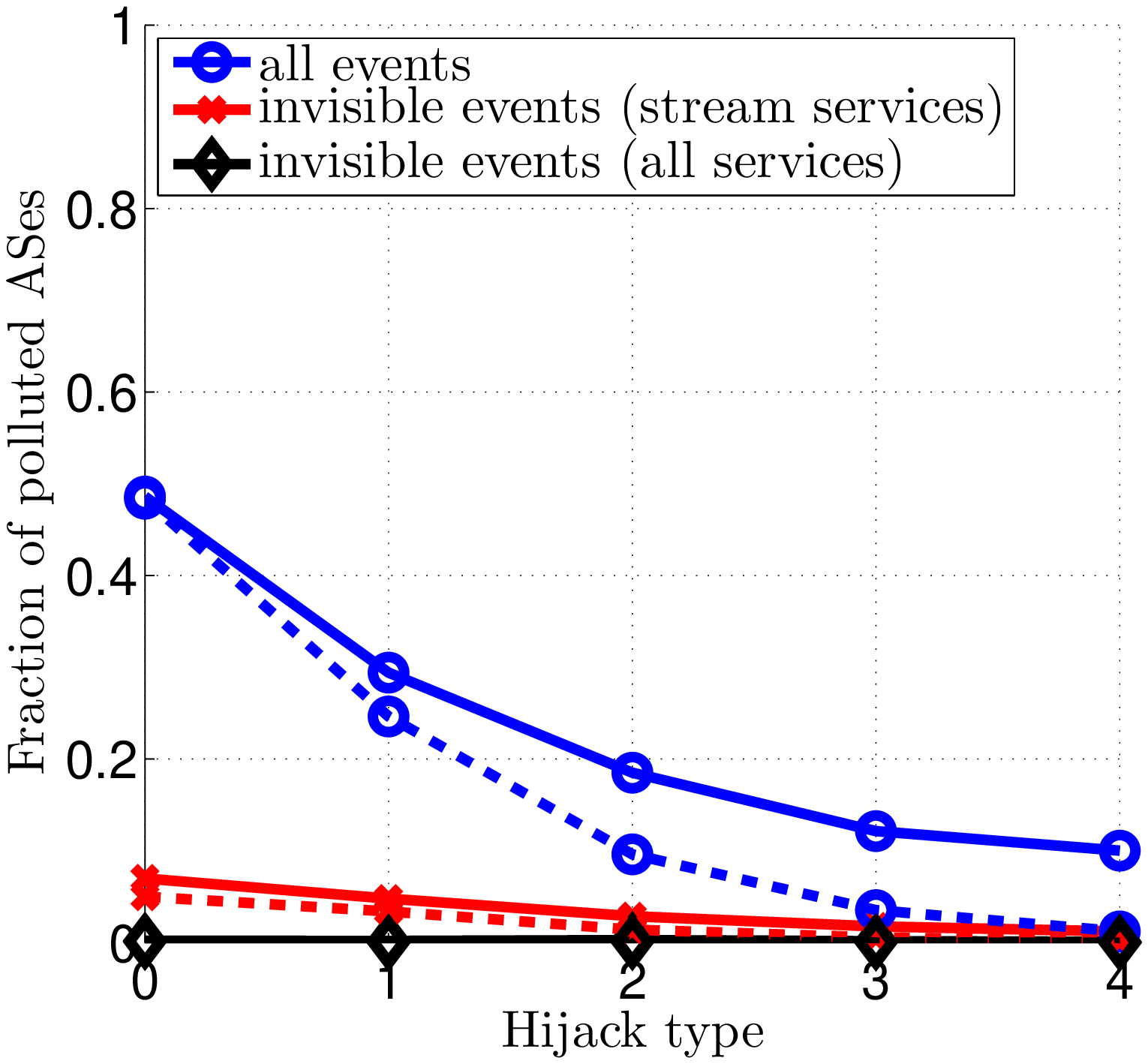}\label{fig:comparison-impact-visibility}}
\caption{\textit{Impact} of different hijack types: (a) CDFs, and (b)
mean (continuous lines) and median (dashed lines) values of the fraction of polluted ASes over 1000 simulations for different hijack
types. %
Hijacking events of all types can have a large impact, with smaller types being on average more impactful.}
\end{figure}

\myitem{Hijacking events of smaller AS-path type tend to have larger impact.} 
Fig.~\ref{fig:hijack-type-impact} shows the \blue{Cumulative Distribution Function (CDF)} of
the percentage of polluted ASes in our simulations.
% over 1000 simulation runs\footnote{add footnote 1000sims are robust (small confidence intervals).}, with varying \{legitimate-AS, hijacker-AS\} pairs. 
The farther the position of the hijacker in the announced path (\ie
as the hijack type increases from 0 to 4), the lower the probability
that a hijack can affect a large fraction of the Internet.
%\ie the CDF curve in Fig.~\ref{fig:hijack-type-impact} for type $m$
%is above the curve for type $n$ when $m>n$.
For hijack types larger than Type-2, in the majority of the cases ($>50\%$)
%($<0.5$ in y-axis) 
their impact is very limited or negligible (\eg
$4\%$ and $1\%$ for Type-3 and Type-4, respectively).%, in contrast to lower types. 

\myitem{All types of hijacks can have a large impact.} 
Comparing the \textit{mean} to the \textit{median} values in Fig.~\ref{fig:comparison-impact-visibility} (blue curves; circle markers) highlights that even with 
Type-4 hijacks there are events with a large (\ie $> 80\%$, see Fig.~\ref{fig:hijack-type-impact}) impact. We
verified that these corner cases happen not because the hijacker AS
has high connectivity, but because of the reciprocal location of the
hijacker and victim ASes in the AS-graph and the respective relationships with
their neighbors.
%Since it is difficult to identify the ASes 
\blue{Hence, network connectivity metrics
alone~\cite{Lad2-Understanding-resiliency-hijacks-DSN-2007}, cannot
always (\ie for all attack types) indicate the potential impact (in
terms of Internet pollution) of an
attacking AS. Since it is difficult to identify the ASes
\footnote{\blue{Ballani \emph{et al.}~\cite{ballani2007study} use simulation to estimate the
probability of impact of hijacking attacks against different ASes in the AS graph.
They show that besides ASes high in the routing hierarchy, even small ASes can hijack and intercept traffic from a non-negligible fraction of ASes, making identification of attackers challenging.}}
 that are capable of launching impactful hijacking attacks (\eg using the methodology of~\cite{qiu2009locating} would require to consider all possible hijacker ASes and attack types), \textit{an operator should be
able to defend their networks against every type of hijacking event}.}

\subsection{Visibility of Hijacks on the Control Plane}
\label{sec:visibility-of-hijacks}
Here we study to which extent different types of hijacks are visible
by monitors of publicly accessible BGP monitoring infrastructure.
Detecting a hijacking event through control-plane monitoring requires
the illegitimate path to propagate to at least one monitor.
Moreover, the more monitors receive such a route, the faster and more
robust (\eg against monitor failures) the detection of a hijack
is.

\begin{figure}
\centering
\subfigure[all monitors]{\includegraphics[width = 0.49\linewidth]{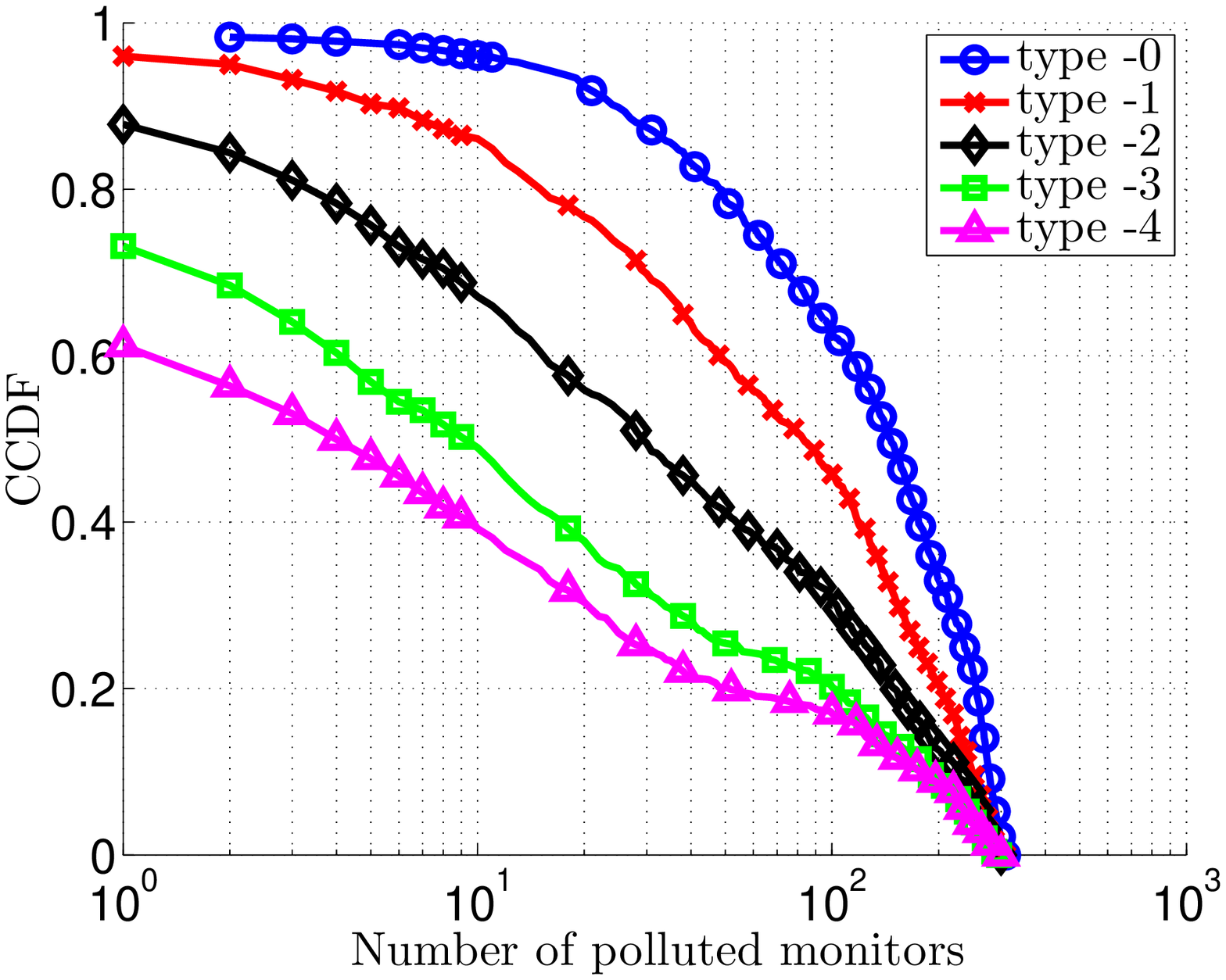}\label{fig:hijack-type-detectability-all}}
\subfigure[streaming monitors]{\includegraphics[width = 0.49\linewidth]{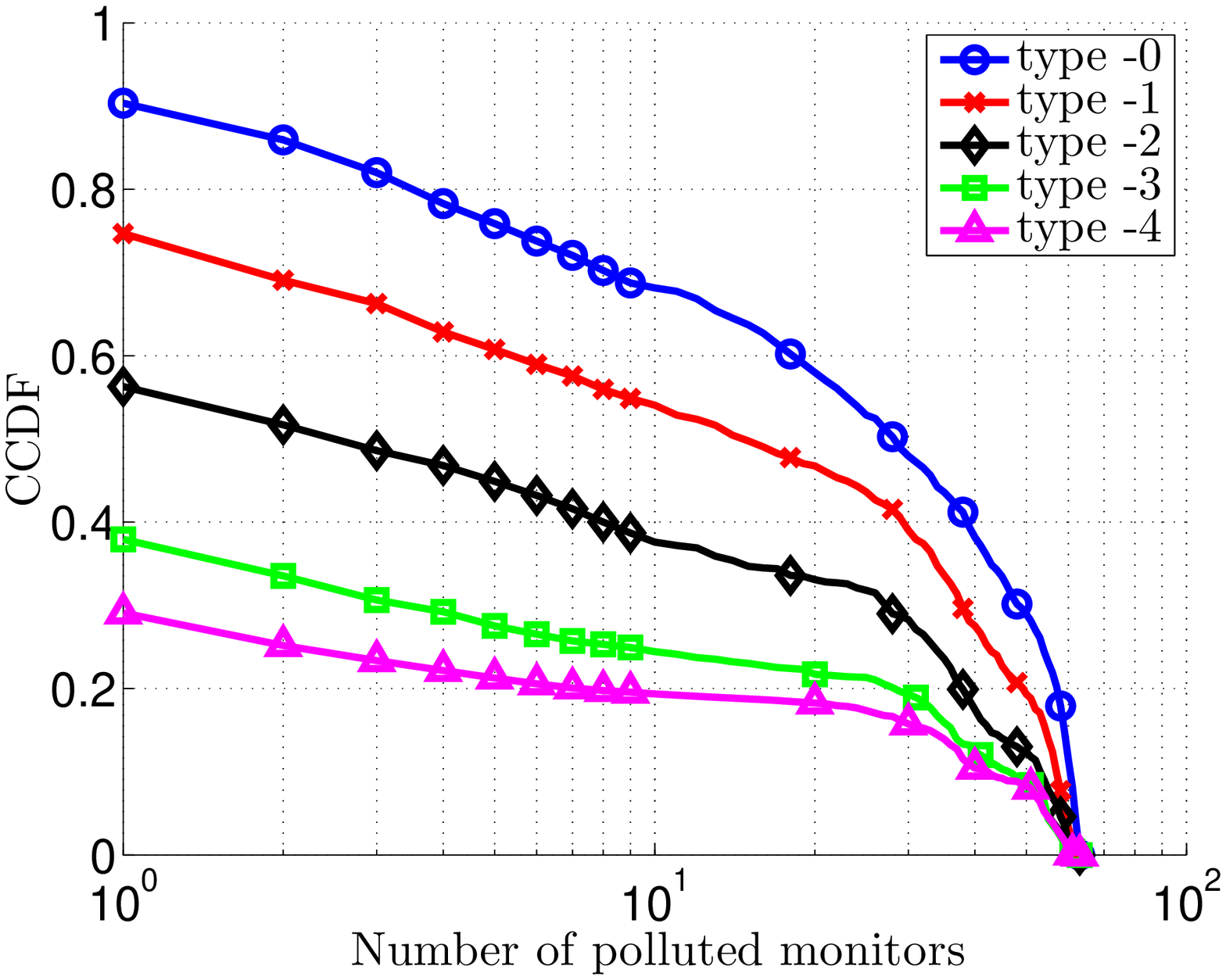}\label{fig:hijack-type-detectability-stream}}
\caption{\textit{Visibility} of different hijack types: CCDFs of the
number of monitors %(x-axis in log-scale) 
that observe an illegitimate route over
1000 simulations for different types, using (a)
\textit{all} and (b) \textit{streaming} monitoring services.
Hijacking events of smaller type are visible with higher probability and to more monitors.}
\label{fig:hijack-type-detectability}
\end{figure}

\myitem{Hijacking events of smaller AS-path type are more visible.}
Fig.~\ref{fig:hijack-type-detectability} shows the distribution
of the number of monitors, from {\em(a)} all monitoring services,
and from {\em(b)} only RIPE RIS and BGPmon streaming services,
that receive an illegitimate path. 
%If an illegitimate path is received by at least one monitor, then the hijacking is visible and its detection through control plane monitoring becomes possible.
%Moreover, the more monitors receive a bogus route, the faster and more robust the detection of a hijack is.
As expected, hijacking events of smaller type are visible with
higher probability and to more monitors (on average), since their
impact on the Internet is larger (see Fig.~\ref{fig:comparison-impact-visibility}). 
%\ed{Any other points we need to stress here? --pavlos}
%\footnote{The streaming services of RIPE RIS and BGPmon receive updates (as seen from our logs) from \red{56} and \red{7} peers (i.e. ASes), respectively. \red{add data for all services}}
Table~\ref{table:average-detectability} gives the percentage of
hijacking events that are \textit{invisible} to the different services (\ie
they do not pollute any of the monitors in our simulations). We can
see that almost all origin-AS hijacks (Type-0) are visible, whereas
hijacks of types 1, 2, 3, and 4 have a higher
probability to remain unnoticed, \eg more than $20\%$ of Type-3
hijacks are not visible by any service.
We also find that the combination of different services always leads
to increased visibility.

\begin{comment}
Comparing the different streaming services, RIPE RIS can observe more
hijacks than BGPmon.  However, BGPmon can observe hijack events that
are invisible by RIPE RIS monitors, and using both services
simultaneously we can observe $1.5\%$ to $5\%$ (for different hijacking
types) extra hijack events than RIPE RIS alone. The visibility from
the two non-streaming services (RouteViews and RIPE RIS) is similar,
however, their combination leads always to increased coverage and visibility.
\end{comment}

\begin{table}[!b]
\begin{minipage}{\linewidth}
\centering
\caption{Percentage of \textit{invisible} hijacking events.
Hijacks of higher types tend to pollute a smaller portion of the
Internet.
%are more probable to remain unnoticed, 
Combining monitoring services always increases visibility.}
%}
%\begin{small}
\begin{tabular}{|c|ccccc|}
\hline
{}									&\multicolumn{5}{c|}{Hijack type}\\
{}						&{0}	 	&	{1} 	&	{2} 	&	{3}	&{4}\\
\hline
{BGPmon (stream)}					&{10.9\%} 		&	{31.6\%} 		&	{53.6\%} 		&	{65.9\%}	&{76.1\%}\\
{RIPE RIS (stream)}					&{~7.1\%}		&	{20.6\%} 		&	{36.7\%} 		&	{50.5\%}	&{63.8\%}\\
\hline
\rowcolor{Gray}
{All stream services}				&{~4.2\%}	 	&	{15.6\%} 		&	{33.1\%} 		&	{47.8\%}	&{62.2\%}\\
\hline
\multicolumn{6}{c}{}\\
\hline
{RouteViews}							&{~1.5\%} 		&	{~4.3\%} 		&	{11.1\%} 		&	{26.5\%}	&{38.0\%}\\
{RIPE RIS}							&{~1.8\%}		&	{~4.0\%} 		&	{13.8\%} 		&	{26.4\%}	&{40.9\%}\\
\hline
\rowcolor{Gray}
{All services}						&{~1.4\%}	 	&	{~3.0\%} 		&	{~9.0\%} 		&	{21.3\%}	&{34.4\%}\\
\hline
\end{tabular}
%\end{small}
\label{table:average-detectability}
\end{minipage}
\end{table}

%As Fig.~\ref{fig:hijack-type-detectability} and
%Table~\ref{table:average-detectability} show, the number of hijacking
%events that are invisible increases with the hijacking type. This is
%due to the fact that the their impact is lower (see
%Sec.~\ref{sec:impact-of-hijacks}) due to the longer routes originated
%by the hijacker. In the following section we investigate jointly the
%impact and visibility of different hijack types, in order to
%understand what are the limits of a hijacking defense system based on
%control-plane monitoring.

%\subsection{Impact of Invisible Hijacking Events}
%A hijacking event is invisible to the control plane monitoring service, if it does not pollute any of the peers/ASes providing feed to the monitoring infrastructure, e.g., if the route collectors does not receive any route containing the hijacker's ASN.

\begin{figure}
\centering
\subfigure[all monitors]{\includegraphics[width = 0.49\linewidth]{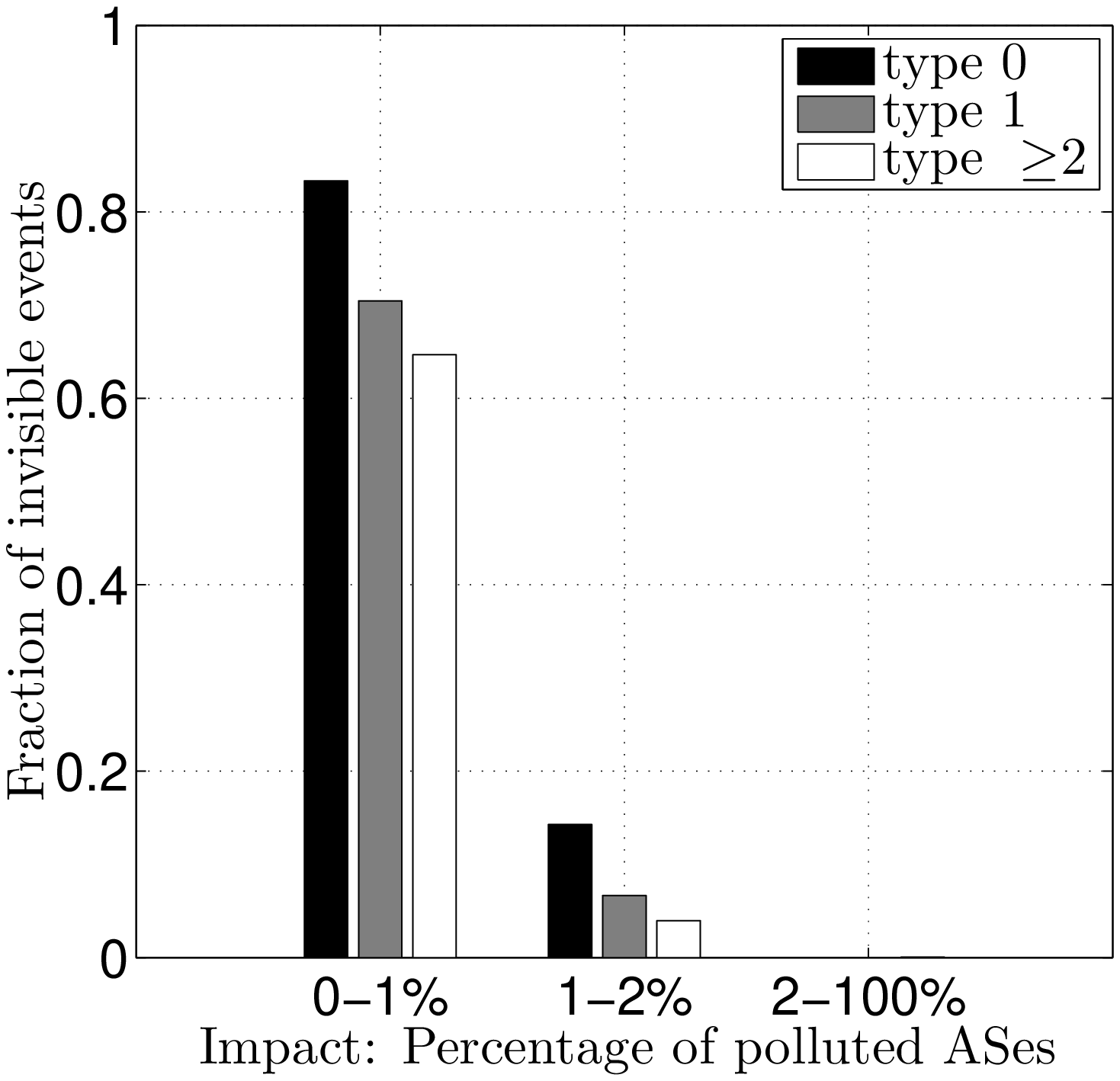}\label{fig:joint-impact-visibility-bars-all}}
\subfigure[streaming monitors]{\includegraphics[width = 0.49\linewidth]{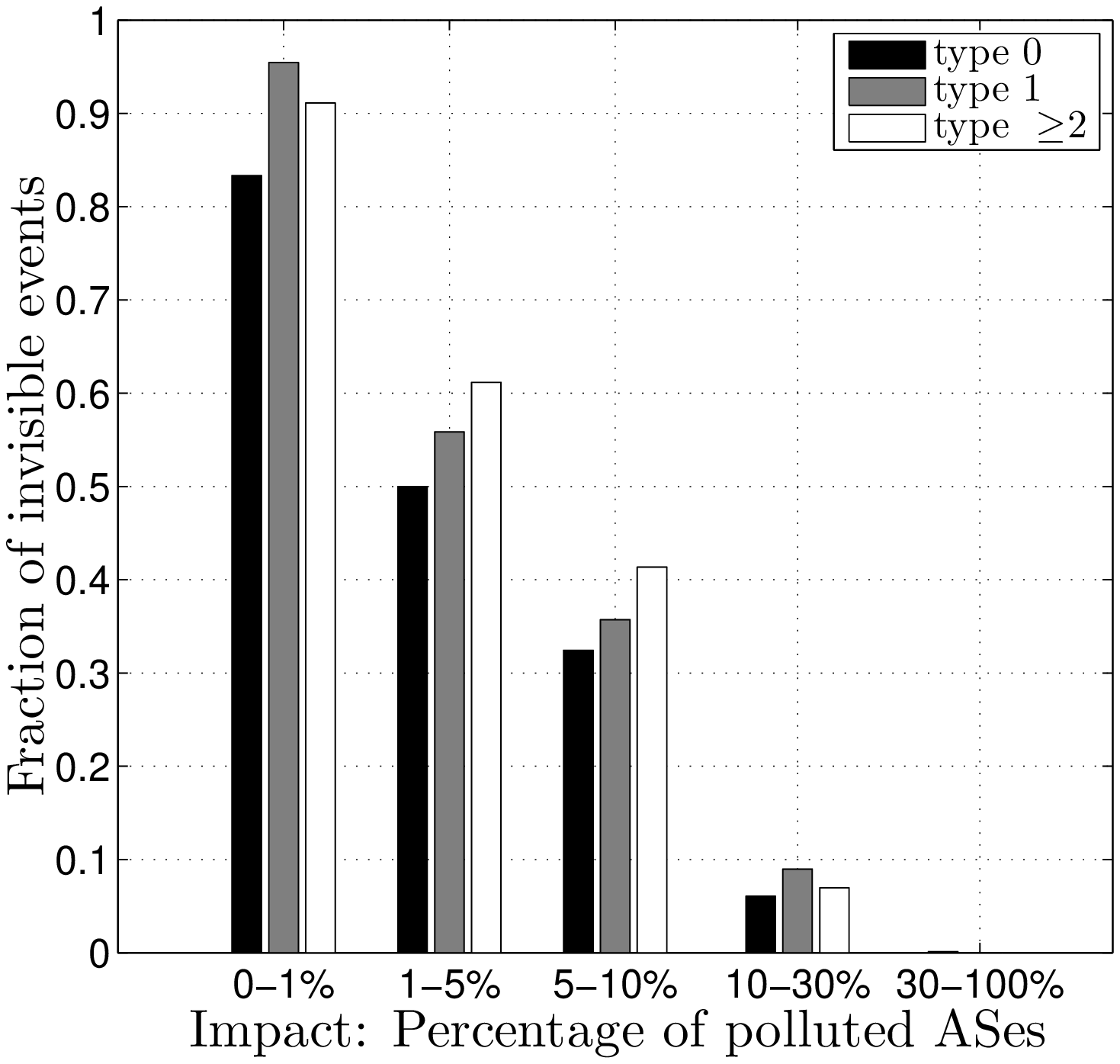}\label{fig:joint-impact-visibility-bars-stream}}
\caption{Fraction (y-axis) of the hijacking events, grouped by
impact (x-axis), that are invisible to (a) all monitoring services,
and (b) streaming monitoring services, for different hijack types
(denoted as bars of different colors). \textit{Note the differences in
the (a)/(b) x-axis.}
Existing monitoring infrastructure can always observe hijacking events
of significant impact.}
% on the control-plane.} %@saved 1 line
\label{fig:joint-impact-visibility-bars}
\end{figure}

\myitem{Hijacking events (of every type) with significant impact are
always visible to monitoring services.}
Fig.~\ref{fig:joint-impact-visibility-bars} shows the fraction
of hijacking events, grouped by their impact, that are invisible to
monitoring services.
Hijacking events that pollute more than $2\%$ of
the Internet are --in our simulations-- always visible to the
monitoring services (Fig.~\ref{fig:joint-impact-visibility-bars-all}),
and the vast majority (\eg more than $85\%$ type-0 hijacks) of those with impact between 1\% and 2\%
are also observed.
The visibility is low only for events with impact less than $1\%$ when
considering all monitors. In total, the mean (median) impact of invisible events is less than $0.2\%$ ($0.1\%$) as shown in Fig.~\ref{fig:comparison-impact-visibility}. These results suggest that existing
infrastructure has already a great potential to enable live
detection of significant hijacking events.  
%Current streaming services (consisting of less monitors; see Table~\ref{table:monitoring-services}) are less efficient; however, they have full visibility for events with impact greater than $30\%$ (Fig.~\ref{fig:joint-impact-visibility-bars-stream}). Comparing Fig.~\ref{fig:joint-impact-visibility-bars-all} and Fig.~\ref{fig:joint-impact-visibility-bars-stream} highlights the expected benefit of RIPE RIS and Route Views accelerating transition to live streaming~\cite{ripe-ris-goes-streaming,bgp-hackathon-2016}.
We find instead that current streaming services have full visibility only for events with impact greater than $30\%$
(Fig.~\ref{fig:joint-impact-visibility-bars-stream}), highlighting 
the potential benefit from RIPE RIS and RouteViews accelerating their
transition to live streaming~\cite{ripe-ris-goes-streaming,bgp-hackathon-2016}.

%\blue{Moreover, we note that selective (e.g., region-specific) BGP announcements --using e.g., BGP communities-- constrain hijacking events within certain regions, while keeping others unaffected, thus making it difficult to detect the incident on a larger scale. However, the large number of available monitors make it very hard for an event to get unnoticed. When the event achieves to be unnoticed by the monitors, this implies that its impact will be very small. In case the hijacker manipulates paths that are legal in one part of the world, so that they are illegally preferred in others too (leading traffic to its AS), ARTEMIS can still detect the associated hijacking event since the ground truth that is supplied by the operator (see \S~\ref{sec:detection}) may take into account region-specific policies (e.g., ``prefix X is advertised by ASX to ASY from the European PoP, but is advertised to ASZ only from the African PoP'').}
\blue{These findings deliver a promising message for using public BGP
monitoring infrastructure to detect hijacks: while a hijacker can
employ several means to achieve a stealthy hijack (\eg to launch Type-N attacks of large $N$, or append BGP communities to limit its visibility within specific regions), the attack can only be invisible at the cost of limited impact.}

\section{Detection Methodology}
\label{sec:detection}

\subsection{Overview} % of Detection Approach}
\label{sec:artemis-overview}

%\ed{I shrunk the text here. The old text is commented in the tex. However, I think that some parts of the old text could be moved to Intro and/or Conclusions (not sure that are mentioned somewhere else in the paper). --pavlos}

%@below stressed in intro
%\myitem{Requirements.} An efficient and practical detection system needs to be \ione \textit{comprehensive} \ie able to detect any type of hijacking attacks, and \itwo \textit{accurate}, \ie zero/low rate of false positives. 
%@we already stress below in the introduction. btw, below should be
%updated to include analysis of literature and its table.
% These requirements are highlighted by the operators in our survey, and are missing from existing approaches (see Sec.~\ref{sec:motivation}; \eg RPKI has little benefits and third-party services return many false alerts).

%\myitem{Approach Overview and Achieved Goals.}
\artemis is run locally by a network and enables a self-operated (\ie not involving third parties) detection of hijacking events for its own prefixes. \artemis \textit{(a)} uses a \textit{local
configuration \blue{file}} with information about the prefixes owned by the
network, and \textit{(b)} receives as input the \textit{stream of BGP updates}
from the publicly available monitoring services and the local routers of the network that operates it.
Comparing the prefix and \texttt{AS-PATH}
fields in the BGP updates with the information in the local
configuration \blue{file}, \artemis can detect any class of hijacking event,
%(\textit{comprehensiveness}), 
and generate alerts.
% only after validating an event (\textit{accuracy}). 

%\blue{\myitem{\textbf{Populating Local Configuration.}}}
%\blue{Local configuration, supplied by the network operator, comprises lists of owned prefixes, legitimate origin ASNs (MOAS), verified AS adjacencies and other information relevant to the detection processes of ARTEMIS. Manually providing the needed information might become a heavy task for the operators of large-scale networks. To facilitate the operation of ARTEMIS, mechanisms that populate (and update) such information automatically can be used. For example, ARTEMIS can communicate with the BGP routers of the network (with iBGP, ExaBGP~\cite{exabgp}, or route reflectors). In our informal conversations with operators of large networks, they have been positive that such an approach is feasible, since it resembles existing methods they already use. Note that while using IRR databases, such as RADb~\cite{radb}, could be an option, these sources may include stale information or miss critical AS-internal information for various reasons (including privacy). Moreover, the update of these databases might not be instantaneous, resulting in false alerts.} 
\blue{The local configuration file is populated by the network
operator, and includes lists of owned ASNs and prefixes, ASNs of
neighboring ASes, and routing policies (\eg ``prefix $p$ is announced
with origin $AS_O$ to neighbors $AS_{n1}$ and $AS_{n2}$''). For ease of use, the local configuration file can be populated (and updated) automatically; for instance, ARTEMIS can communicate with the BGP routers of the network (with iBGP, ExaBGP~\cite{exabgp}, or route reflectors).}

\myitem{\blue{False Positives (FP) and Negatives (FN).}} Table~\ref{table:taxonomy-detection} summarizes the FP--FN performance
of the different detection criteria used in our approach for each attack
scenario (discussed in  
\S~\ref{sec:detection-types-other},
\S~\ref{sec:detection-types-0-1},
\S~\ref{sec:detection-types-N}).
%which we illustrate in the following sections. 
By default, our approach does not introduce FN for any attack
scenario. The only possible FN are
the events not \textit{visible}%\footnote{Which would become visible by deploying additional monitors.}
~by the monitoring infrastructure (\S~\ref{sec:visibility-of-hijacks}),
which have very limited impact on the control plane
(Figs.~\ref{fig:comparison-impact-visibility} and~\ref{fig:joint-impact-visibility-bars}).
We generate potential FP (at a very low rate)
%: $<0.3$ alerts per day estimated for the vast majority of the ASes in our experiments) 
only for exact-prefix hijacking events of Type-N, $N\geq2$; however, for the detection of this class of
events, \artemis optionally allows the operator to trade \ione speed
for increased accuracy, and \itwo potential FN related to
events with negligible visible impact (\eg seen by only 1 monitor) for less FP.

\begin{table*}
\centering
\caption{Detection of the different BGP prefix hijacking attacks by \artemis.}
\label{table:taxonomy-detection}
\begin{tabular}{|ccc|ccccc|}
\hline
%{Prefix}		&	{AS-PATH (Type)		&	{Data-Plane}		& {False Positives} & {False Negatives}	& {Detection Approach}} 
\rowcolor{Gray}
\multicolumn{3}{|c|}{Hijacking Attack} & \multicolumn{5}{c|}{\artemis Detection}\\
\hline
\rowcolor{Gray}
{Prefix}		&	{AS-PATH}		&	{Data}		& {False} 								& {False}				&{\blue{Detection}} &{\blue{Needed Local}} & {Detection}\\
\rowcolor{Gray}
{}			&	{(Type)}			&	{Plane}		& {Positives (FP)} 						& {Negatives (FN)}		&{\blue{Rule}} 		&{\blue{Information}} & {Approach}\\
\hline
{Sub-prefix}&	{*}				&	{*}			& {None}									& {None}					&{\blue{Config. vs BGP updates}}&{\blue{Pfx.}}& {Sec.~\ref{sec:detection-types-other}}\\
\hline
{Squatting}&		{*}				&	{*}			& {None}									& {None}					&{\blue{Config. vs BGP updates}}&{\blue{Pfx.}}& {Sec.~\ref{sec:detection-types-other}}\\
\hline
{Exact}&			{0/1}			&	{*}			& {None}									& {None}					&{\blue{Config. vs BGP updates}}&{\blue{Pfx. + ASN}}& {Sec.~\ref{sec:detection-types-0-1}}\\
{}&			{}			&	{}			& {}									& {}					&{}&{\blue{(+ neighbor ASN)} }& {}\\
\hline
{Exact}&			{$\geq 2$}		&	{*}			& {$<0.3$/day for}						& {None}					&{\blue{Past Data vs BGP updates}}&{\blue{Pfx.+ Past AS links}} & {Sec.~\ref{sec:detection-types-N}}\\
{}&				{}				&   {}			& {$>73\%$ of ASes}						& {}						&{\blue{(bidirectional link)}}&{}& {\textit{Stage 1}}\\
\hline
{Exact}&			{$\geq 2$}		&	{*}			& {None for $63\%$ of ASes}				& {$<4\%$}				&{\blue{BGP updates }}&{\blue{Pfx.}}& {Sec.~\ref{sec:detection-types-N}}\\
{}&				{}				&   {}			& {($T_{s2}=5min$, }	&{}					&{\blue{(waiting interval,}}&{} &{\textit{Stage 2}}\\
{}&				{}				&   {}			& {$th_{s2} >1$ monitors)}	&{}			&{\blue{bidirectional link)}}&{} &{}\\
\hline
\end{tabular}
\end{table*}

\subsection{Detecting Sub-prefix Hijacks \& Squatting }
\label{sec:detection-types-other}

Sub-prefix hijacks are the most dangerous, since they can pollute the
entire Internet due to the longest prefix matching employed by the BGP
decision process.
They are also among the most problematic when using third-party
services, since each time an AS decides to announce a longer prefix
or to de-aggregate a prefix, it either needs to communicate this
information in advance to the third-party service or it will receive a
false-positive alert from it. For this reason, often sub-prefix
detection is not even implemented/enabled (\S~\ref{sec:motivation}).

\myitem{\artemis returns 0 false positives and 0 false negatives 
for \textit{all} sub-prefix hijacking events --- independently of
the Type being 0, 1, 2, ... .} To detect these events,
the network operator stores in the \textit{local configuration \blue{file}} of
\artemis an up-to-date \textit{list of all owned and announced
prefixes}. When a sub-prefix hijack takes place, the monitoring
services observe BGP updates for this sub-prefix (the entire
Internet is polluted), and \artemis immediately detects it, \blue{since the sub-prefix is not included in the list of announced prefixes}.
%these
%events %without any false positives/negatives, since it has full
%knowledge of which prefixes are legitimately announced.
%
Such a detection becomes trivial with our approach (\ie leveraging local information). However, this is an important result: %Even if such a detection methodology is trivial, this is an important result: 
without this detection in place, attackers can remain stealthy
by announcing a sub-prefix, which allows them to avoid announcing an
illegitimate \texttt{AS-PATH} (and can further increase stealthiness by
carrying the attack on the data plane as a Man-in-the-Middle
\cite{defcon16-attack}).
In the following sections we illustrate how \artemis detects the
remaining classes of attacks when \textit{exact-prefix} hijacking is
involved instead.

%However, \artemis immediately detects these events without
%any false positives/negatives, since it has full knowledge of which
%prefixes are legitimately announced (stored in the local configuration).

%The real-time
%services~\cite{bgpmon,ripe-ris-real-time} provide the option to
%monitor all sub-prefixes of a prefix. If a BGP update for a subprefix
%of an announced prefix is received, \artemis immediately generates a
%subprefix hijack alert.
%The approaches of Sections~\ref{sec:detection-types-0-1} and
%\ref{sec:detection-types-N} apply also for subprefix hijacking
%events, depending on the type of the event (0,1,$N\geq2$).
\myitem{\artemis returns 0 false positives and 0 false negatives 
for \textit{all} BGP squatting events.} Checking
against the operator's list of actually announced prefixes, has the added benefit of detecting \textit{BGP
squatting} as well; a technique commonly used by
spammers, in which a (malicious) AS announces space owned but not
announced by another AS~\cite{nanog-squatting,schlamp2013forensic}.
%\artemis detects illegitimate announcements for owned unannounced prefixes.
%\textbf{Squatting.} The \{origin ASN, neighbor ASN\} fields for
%unannounced prefixes are empty (see
%Section~\ref{sec:detection-types-0-1}). Hence, \artemis immediately
%(\ie without any need for further processing) generates an alert
%upon the reception of a BGP update for a prefix that matches
%unannounced owned space.
%
%In the following sections, we discuss detection criteria for
%\textit{exact prefix hijacking}.

\subsection{Detecting Type-0/1 Exact Prefix Hijacks}
\label{sec:detection-types-0-1}

%To detect these events, the network operator stores in the \textit{local configuration}
%an up-to-date list of all owned, announced%\footnote{The lists are empty for owned but unannounced prefixes.} 
%~prefixes, as well as the
%following information \textit{per prefix}:

The network operator provides also in the \textit{local configuration file} the following information \textit{per prefix}:

%\red{This information can also be generated and updated automatically
%by periodically parsing router configuration files.}
%@ not entirely sure about how solid this is. \eg what are other
%ASes authorized to originate is not in the router conf file. We may
%say internal conf files but that defeats the point of emphasizing.
%also not sure it belongs here and if it clashes with the mitigation
%configuration

\setlist{nolistsep}
\begin{itemize}[leftmargin=*]

\item \textit{Origin ASN(s)}: the ASNs authorized to originate the
prefix.

\item \textit{Neighbor ASN(s)}: the ASNs 
with which there are direct BGP sessions established, where the
prefix is announced.
%\blues{; this list indicates the routing policy for each owned prefix}.
%\footnote{When a prefix is announced from several origin ASes (MOAS), the list of neighbors can be provided per \{prefix, origin ASN\} tuple.}

\end{itemize}

\noindent
For every BGP update it receives from the
monitors, \artemis extracts
the \texttt{AS-PATH} field, and compares the announced prefix, as well
as the first and second ASNs in the \texttt{AS-PATH}, with the
\{prefix, origin ASN, neighbor ASN\} information in the local file. If
the \texttt{AS-PATH} does not match the information in the local
file, a hijack alert is generated.
% The process of verifying the \texttt{AS-PATH} field in the received
% BGP update against the local list is simple and lightweight, with
% negligible delay (\eg $<1$s in our experiments --
% see~Section~\ref{@@}).
%@@ do we really need above (commented)? --alb

\myitem{\artemis detects all Type-0 and Type-1 hijacks that are
visible to the monitors (\ie 0 false negatives for visible
events).} As in \S~\ref{sec:detection-types-other}, since \artemis
leverages \textit{ground truth} provided by
the operator itself, all illegitimate paths that are
visible by the monitors are always detected as hijacks. 
%As a result, our findings in \S~\ref{sec:understanding-hijacks} on the \textit{visibility} and \textit{impact of invisible} exact-prefix hijacking events apply directly to the \textit{detectability} of hijacking events.  Specifically, the vast majority of these events can be detected (see Table~\ref{table:average-detectability} for Type-0/1 events), while the undetectable events have a very limited impact on the control plane (see Figs.~\ref{fig:comparison-impact-visibility} and~\ref{fig:joint-impact-visibility-bars}).

\myitem{\artemis returns 0 false positives for Type-0/1
hijacking events.} Any BGP update that does not match the local lists
\{prefix, origin ASN, neighbor ASN\}, indicates \textit{with
certainty} an announcement originated illegitimately by another
network (\ie  without the consent of the prefix owner). 
%The lists of \{prefix, origin ASN, neighbor ASN\} are provided by the
%network operator and are treated as ground truth for the associated
%network. Any BGP update that does not match these lists indicates
%thus with certainty an announcement originated illegitimately by
%another network (\ie  without the consent of the network that owns
%the prefix).
%
%On the contrary, detection services operated by third-party
%organizations, \eg~\cite{commercial-bgpmon}, might not have complete
%information of the ground truth (\ie origin-ASNs, and their peering
%policies), and are thus susceptible to incorrect detection alarms.
%Moreover, with \artemis the network \textit{does not disclose any
%information} (\eg routing policies) to third parties, which is
%usually a concern of operators (cf. Section \ref{sec:motivation}).

\subsection{Detecting Type-N, N$\mathbf{\geq}$2, Exact Prefix Hijacks}
\label{sec:detection-types-N}
Detecting Type-N, $N\geq 2,$ hijacking events requires a different
approach than Type-0/1 events, since the operator might not be aware
of all its $2^{nd}$, $3^{rd}$, ... hop neighbors. To this end,
\artemis \ione detects all suspicious Type-N, $N\geq 2$, events, \ie
when new links\footnote{We consider only new links and not policy
violations on existing links (as, \eg \cite{Shi-Argus-IMC-2012}\blue{\cite{qiu2007detecting}}),
since routing policies are not publicly available, and inferences
based on existing datasets would lead to a very high number of false
alerts; \eg~\cite{routing-policies-in-the-wild} shows that around $30\%$ of the observed routes are not in agreement with the available routing policy datasets.} appear in routes towards the operator's prefixes, \itwo filters out as many legitimate events as possible, 
%reaching a certain confidence level per event, 
and \ithree augments alerts with information about the estimated impact of the remaining suspicious events.

\begin{comment}
%%%%%%%%%%   ALBERTO's VERSION   %%%%%%%%%%
Differently from Type-0 and Type-1 events, for which 
%---by leveraging the operator's knowledge --  
\artemis achieves 
%detection with no false positives, 
0\% false positive rate,
detecting  Type-N, $N\geq 2,$ hijacking events may yield to
false alerts, since the operator might not be aware of all its
$2^{nd}$, $3^{rd}$,... hop neighbors.
Therefore, for Type-N, $N\geq 2,$ events, \artemis \ione detects all potential
hijacks (\ie when new links appear in routes towards the operator's
prefixes), \itwo filters out as many legitimate events as possible,
and \ithree augments alerts with information about the estimated
impact of the suspicious event.
%%%%%%%%%%   end of ALBERTO's VERSION   %%%%%%%%%%
\end{comment}

Specifically, \artemis uses a configurable two-stage detection approach, where the
operator can trade detection speed (\textit{Stage 1}) for increased accuracy and impact
estimation (\textit{Stage 2}).~\textit{Stage 1} detects all potential hijacking events as soon as
they are observed by a monitor (\ie typically with few seconds
latency), filters out benign events based on information that is
available at detection time, and generates alerts for suspicious events. An optional \textit{Stage 2} collects additional
information within a (configurable) time window $T_{s2}$ following the detection from \textit{Stage 1}, in order to
(a) increase the chance of filtering out a benign event, and (b) provide 
the operator with an estimate of the impact of the event in case it is
still recognized as suspicious.

\begin{comment}
%%%%%%%%%%   ALBERTO's VERSION   %%%%%%%%%%
To this end, \artemis uses a configurable two-stage detection approach, where the
operator can trade detection speed (\textit{Stage 1}) for increased accuracy and impact
estimation (\textit{Stage 2}).
%
\textit{Stage 1} detects all potential hijacking events as soon as
they are observed by a monitor (\ie typically with few seconds
latency), filters out benign events based on information that is
available at detection time, and generates alerts for suspicious events. An optional \textit{Stage 2} collects additional
information in the 5 minutes following the
detection from the previous stage\footnote{The 5 minute value can actually
be configured by the operator, but in the paper ....}, in order to
increase the chance of filtering out a benign event and to provide 
the operator with an estimate of the impact of the event in case it is
still recognized as suspicious.
%%%%%%%%%%   end of ALBERTO's VERSION   %%%%%%%%%%
\end{comment}

\subsubsection{Stage 1}%\myitem{Stage 1}. 
For Type-N, $N\geq 2,$ detection, \artemis stores locally the following lists
of \textit{directed} AS-links \blue{(with related metadata)}:

\begin{itemize}[leftmargin=*]
\item \textit{previously verified AS-links list}: all the AS-links that appear in a path towards an
owned prefix and have been verified by \artemis in the past.
%Once \artemis has verified an AS-link appearing in a path towards an
%owned prefix, it stores it in the local configuration and avoids verifying it again. 

\item \textit{AS-links list from monitors}: all the AS-links
in the AS-path towards \textit{any} prefix (\ie owned by any AS)
observed by the monitors, in a sliding window of the last 10 months.
This list represents an historical view of observed (directed) AS-links.
The 10-month time frame should
accommodate the observation of most of the backup routes
\cite{chen2009sidewalk}.
%which is created \textit{offline}. The list is collected using the
%BGPstream tool, by downloading all BGP messages seen at the monitors
%within a period of \red{X} months. \ed{how is this period selected?
%--pavlos/vassilis}. The list is updated regularly, every \red{Y} time
%units, where \red{Y} is a tunable parameter; small \red{Y} leads to
%fresh, but maybe incomplete information, while large \red{Y} to more
%complete albeit potentially stale information. \ed{how do we select
%the update frequency?  --pavlos/vassilis}

\item 
\textit{AS-links list from local BGP routers}: all the AS-links
observed in the BGP messages received by the BGP routers of the
network operating \artemis. The list is collected by connecting to
the local BGP routers (\eg via ExaBGP~\cite{exabgp} or with BGPStream and BMP
\cite{bgpstream-v2-beta, scudder2016bgp}), and receiving
every BGP update seen at them, or alternatively querying a route
server.  This list is also updated continuously within a 10-month
sliding data window.
\end{itemize}

\noindent The detection algorithm is triggered when a monitor receives a BGP update (for a monitored prefix) whose \texttt{AS-PATH} contains
an N-hop ($N\geq2$) AS-link that is not included in the \textit{previously verified AS-links list}. \blue{Let $AS_{V}$ be the victim AS (operating \artemis), the new AS-link be between ASes $AS_{X}$ and $AS_{Y}$, and the \texttt{AS-PATH} of the BGP update be
\begin{align*}
P^{new}_{(X,Y)} 	&= \{AS_{\ell 1}, AS_{\ell 2}, ..., AS_{X}, AS_{Y}, AS_{r 1}, AS_{r 2}, ..., AS_{V}\}\\
				&= \{\mathcal{L}^{new}, AS_{X}, AS_{Y}, \mathcal{R}^{new}, AS_{V}\}
\end{align*}
where $\mathcal{L}^{new}=\{AS_{\ell 1}, AS_{\ell 2}, ...\}$ denotes the set of ASes appearing in the path after (left of) the suspicious link, and $\mathcal{R}^{new}= \{AS_{r 1}, AS_{r 2}, ...\}$ before (right of) the suspicious link. Note that the type of the attack is $N=2+|\mathcal{R}^{new}|$%, where $|\mathcal{R}^{new}|$ is the number of ASes before the suspicious link
.
}
\blue{The observation of $P^{new}_{(X,Y)}$ is considered as a suspicious event (and previous works would raise an alarm~\cite{Shi-Argus-IMC-2012,qiu2007detecting}). However, it is possible that $P^{new}_{(X,Y)}$ corresponds to a legitimate event (\eg change of a routing policy) that made the link $AS_{X} - AS_{Y}$ visible to a monitor. To decrease the number of false alarms, \artemis applies the following filtering rules.}

\myitem{Rule 1 (bi-directionality).} \blue{Check if the new link $AS_{X} - AS_{Y}$ has been observed in the opposite direction (\ie $AS_{Y} - AS_{X}$) in the \textit{AS-links list from monitors} and/or \textit{AS-links list from local BGP routers}. If the reverse link $AS_{Y} - AS_{X}$ is \textit{not} previously observed, the event is labeled as suspicious.}

\myitem{Rule 2 (left AS intersection).} \blue{Otherwise (\ie the reverse link $AS_{Y} - AS_{X}$ is previously observed), check the AS paths in all the BGP updates containing the reverse link. Let $\mathcal{P}^{old}$ be the set of all these AS-paths, and denote
\[P = \{\mathcal{L}_{P}, AS_{Y}, AS_{X}, \mathcal{R}_{P}\}~~~~~,~~~\forall P \in\mathcal{P}^{old}\]
Then, collect all the sets of ASes $\mathcal{L}_{P}$, $\forall P\in \mathcal{P}^{old}$, that appear after (left of) the reverse link, and calculate the intersection of all these sets, i.e., $\mathcal{L}^{old} = \bigcap _{P\in\mathcal{P}^{old}} \mathcal{L}_{P}$. If $\mathcal{L}^{old}$ is not empty, and at least one AS in $\mathcal{L}^{old}$ appears also in $\mathcal{L}^{new}$ (\ie $\mathcal{L}^{old} \bigcap \mathcal{L}^{new} \neq \emptyset$) in the new received path $P^{new}_{(X,Y)}$, then the event is labeled as suspicious. If $\mathcal{L}^{old} \bigcap \mathcal{L}^{new}=\emptyset$, the event is labeled as legitimate.
%\end{itemize}
}

\blue{\artemis uses these two filtering rules to identify suspicious
announcements of fake links that either contain the attacker's ASN
(\textit{Rule 1}) or do not (\textit{Rule 2}). The rationale behind the two rules is detailed in the following.
}

\blue{
\textit{Rule 1} detects events where the hijacker (\eg $AS_{X}$) is
one end of the fake link. While $AS_{X}$ can fake an
\textit{adjacency} with $AS_{Y}$, and the link $AS_{X} - AS_{Y}$
appears in the polluted routes, the reverse link (\ie $AS_{Y} -
AS_{X}$) is not advertised by $AS_{Y}$ or other networks, and thus not
seen by any monitor. It is impossible for an attacker
\textit{controlling a single AS} to make such a fake link
appear in both directions in order to evade the detection of
\textit{Rule 1}\footnote{The only way for $AS_{X}$ to announce a path
containing $AS_{Y}-AS_{X}$ is to announce a path with a loop (\eg
$\{AS_{X},...,AS_{Y},AS_{X},...\}$), but \artemis detects and discards
announcements with loops instead of adding them to the
\textit{AS-links list from monitors} list.}. Hence, observing an
AS-link $AS_{X} - AS_{Y}$ in both directions, eliminates the
possibility that $AS_{X}$ advertises a fake adjacency. On the
contrary, observing a new link in only one direction cannot guarantee
a legitimate announcement and thus causes \artemis to raise an alert. 
}

\blue{
\textit{Rule 1} can be evaded only if the hijacker \ione \textit{controls at least two ASes}%, and announces both directions of such a fake link for the same or different prefixes, and both announcements are visible to the monitors
, or \itwo announces a fake link not containing its ASN. While the
former case violates our threat model and is out of the scope of the
paper, we apply \textit{Rule 2} to detect the latter case. For instance, a hijacker $AS_{Z}$ can announce to its neighboring ASes two paths containing a fake link $AS_{X}-AS_{Y}$ in both directions:
\begin{align*}
P_{1} &= \{AS_{Z}, ..., AS_{X}, AS_{Y}, ...\}\\
P_{2} &= \{AS_{Z}, ..., AS_{Y}, AS_{X}, ...\}
\end{align*}
However, in its announcements, the hijacker has to append its ASN as the last (leftmost) AS in the path, before further propagation (see \S~\ref{sec:taxonomy-as-path} and RFC4271~\cite{BGPv4}). Hence, in all BGP updates containing the fake link $AS_{X}-AS_{Y}$ in any direction, the AS of the hijacker will appear on the left of the fake link. \textit{Rule 2} identifies whether there exists a common AS in all (new and old) announcements involving any direction of the new (suspicious) link. If at least one AS appears in all paths, then the event is considered suspicious.
}

\myitem{\artemis's \textit{Stage 1} returns 0 false negatives.}
\blue{\artemis detects any illegitimate announcement that is seen by the monitors and
contains a fake link with (\textit{Rule 1}) or without (\textit{Rule
2}) the hijacker ASN at its ends.
It is not possible for an attacker conforming to the threat model of \S~\ref{sec:hijack-types-characteristics} to evade these rules, as long as its announcements are visible. 
}

\begin{figure}
\centering
\subfigure[Effects of Stage 1]{\includegraphics[width=0.49\linewidth , height=0.40\linewidth]{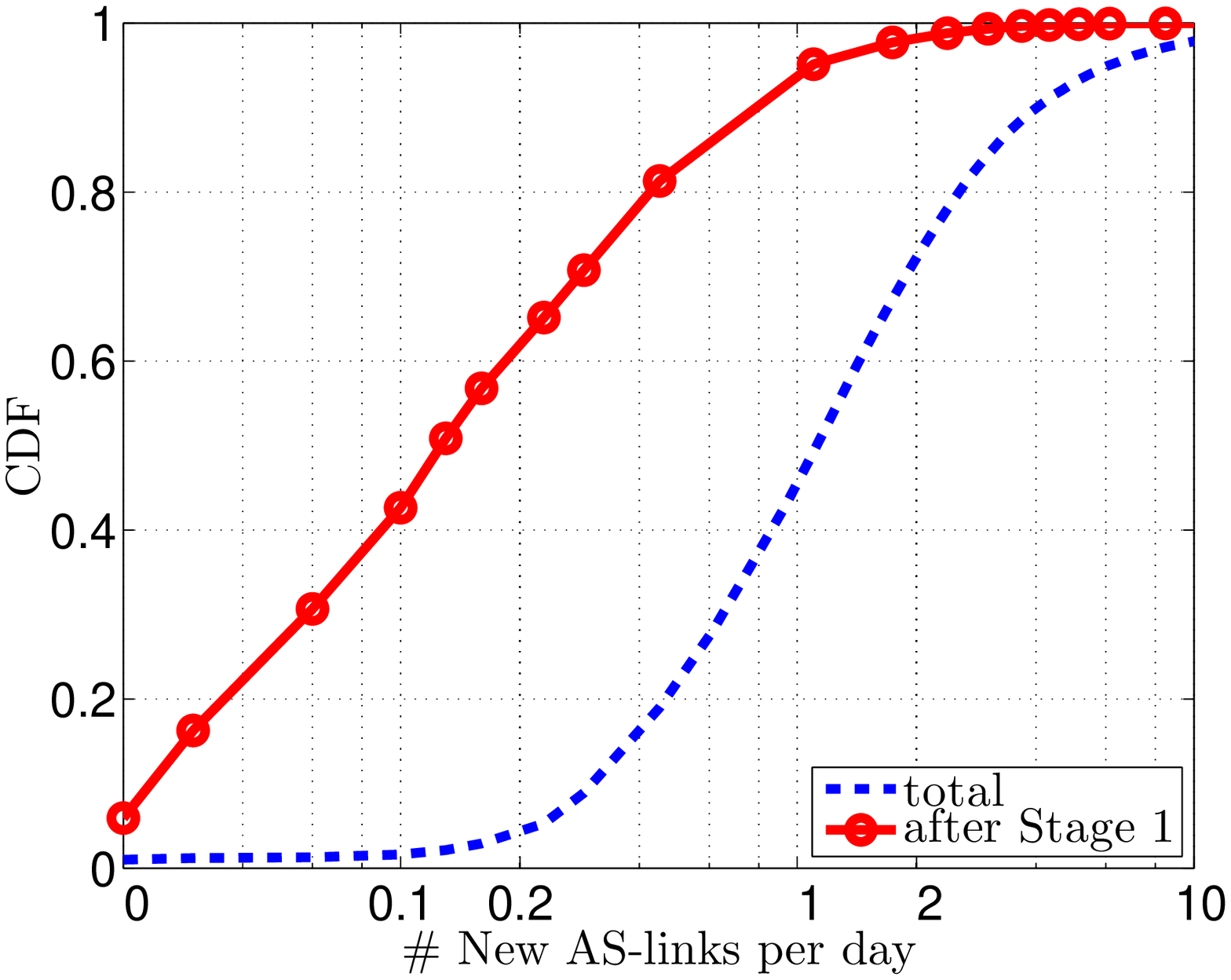}\label{fig:rules}}
%\subfigure[]{\includegraphics[width=0.49\linewidth , height=0.40\linewidth]{./figures/tmpfig1}\label{fig:rules}}
\subfigure[Effects of Stage 2]{\includegraphics[width=0.49\linewidth , height=0.40\linewidth]{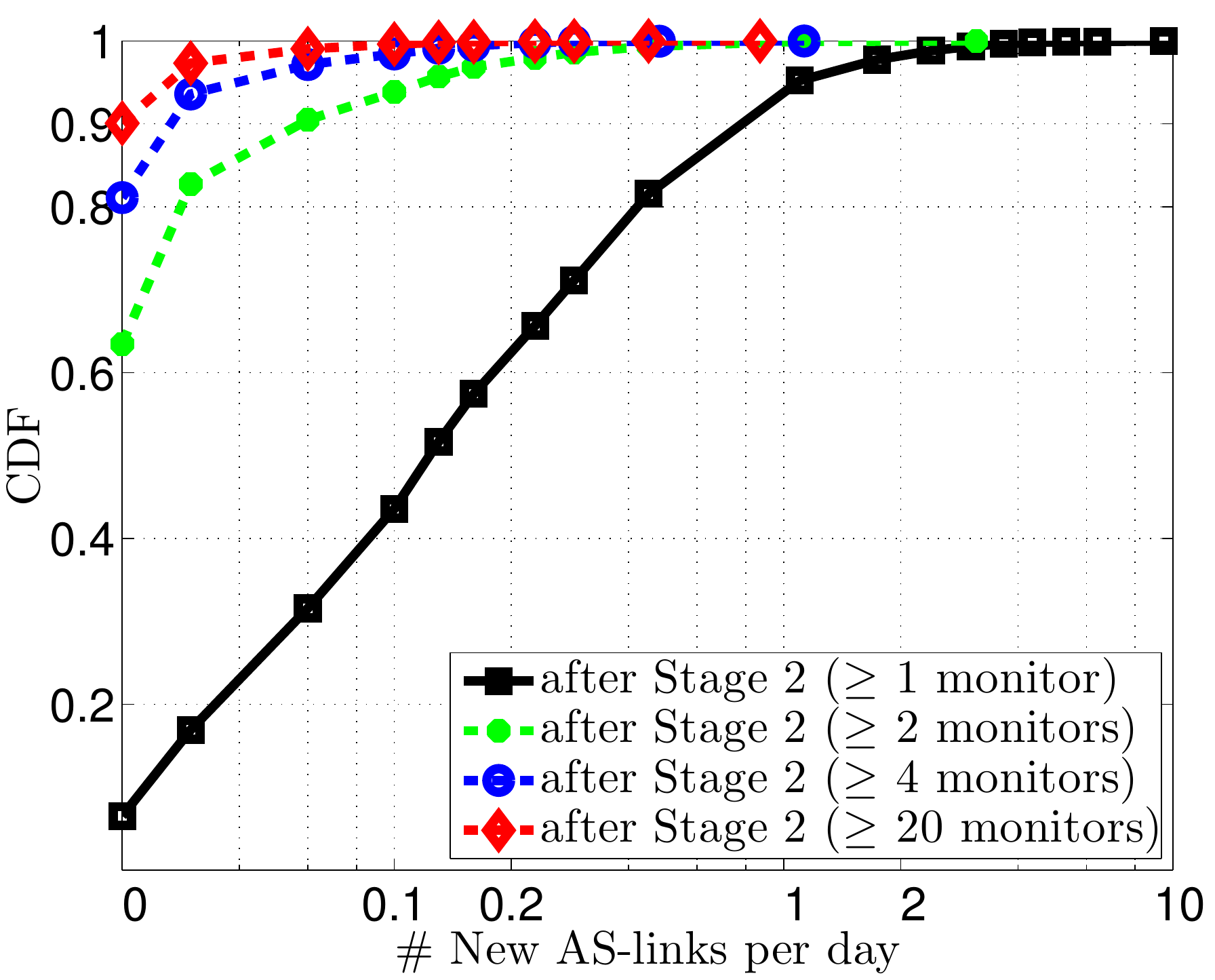}\label{fig:cdf-new-links-times}}
%\subfigure[]{\includegraphics[width=0.49\linewidth]{./figures/tmpfig2}\label{fig:cdf-new-links-times}}
%\subfigure[]{\includegraphics[width=0.49\linewidth]{./figures/fig_cdf_new_links_times_all}\label{fig:cdf-new-links-times}}
%\subfigure[]{\includegraphics[width=0.49\linewidth]{./figures/fig_probe_delay_cdf}\label{fig:cdf-new-links-times}}

\caption{CDF of the number of new AS-links seen at the monitor AS per day, per origin AS: (a) before and after applying Stage 1 - ARTEMIS detection algorithm for Type-N, $N\geq2$, is rarely triggered and \textit{Stage 1} dramatically reduces the number of FP; (b) after applying \textit{Stage 2} ($T_{s2}=5$ min), with different thresholds for the minimum number of monitors that see the suspicious event - requiring at least 2 (or more) monitors to see the event, greatly reduces the number of FP.}
% \textbf{Transiting from 1 ($\geq0\%$ projected impact) to 2 monitors ($\geq0.5\%$ projected impact) greatly reduces the number of false positives}}

%\caption{CDF of the number of new AS-links seen at the monitor AS
%per day, per origin AS: (a) before and after applying Stages 1 and 2 - ARTEMIS detection algorithm for
%Type-N, $N\geq2$, is rarely triggered and \textit{Stage 1}
%dramatically reduces the number of FP, while
%\textit{Stage 2} further reduces this number; (b) after applying \textit{Stage 2} ($T_{s2}=5$ min), with different
%thresholds for the minimum number of monitors that see the suspicious event - requiring at least 2 (or more) monitors to 
%see the event, greatly reduces the number of FP.}
%% \textbf{Transiting from 1 ($\geq0\%$ projected impact) to 2 monitors ($\geq0.5\%$ projected impact) greatly reduces the number of false positives}}
\end{figure}

\myitem{The \artemis detection algorithm for Type-N, $N\geq2$,
hijacks, is rarely triggered.}
%After a period of stable deployment, these events are relatively
%unfrequent:
To understand how often the detection algorithm would be triggered, we
ran our algorithm on 1 month of real BGP data, emulating running \artemis for 
each and every AS announcing prefixes on the Internet. Specifically, we
processed all the BGP updates observed by RIPE RIS and RouteViews
monitors (a total of 438 ASes hosting at least 1 monitor each)
% --\textit{monitor ASes} in the following
 between April 2016 and March
2017. Then, for each AS that originated IPv4 or IPv6 prefixes in March
2017% (\textit{origin AS} in the following)
, we
identified the links appearing for the first time in paths towards
their originated prefixes, during the same month. 
Fig.~\ref{fig:rules} shows the CDF (blue/dashed curve) of the number of new
AS-links an origin AS sees (through the monitor ASes) per day towards
its own prefixes: on average,
within the month of March 2017, $72\%$ of the origin ASes saw less
than $2$ new links per day.
%60\% of the ASes on average did not see any new link, while less than 22\% saw more than one.
%\textbf{\textit{applying \textit{Stage 1} 

\myitem{\textit{Stage 1} dramatically reduces the
number of suspicious events}. We apply the filtering of \textit{Stage
1} to the
previous data; we considered only the \textit{AS-links list from monitors} (since we do not have access to the local routers of all the ASes). Fig.~\ref{fig:rules} shows the CDF of the number of the aforementioned
events that fail \textit{Stage 1} (red/circles curve):
$73\%$ of the origin ASes see less than $1$ suspicious event
every $3$ days.

\subsubsection{Stage 2 (optional)}%\myitem{Stage 2}.
\label{sec:stage2}

Stage 2 introduces an extra delay ($T_{s2}$) in exchange for \ione refined filtering
and \itwo the ability to estimate the impact of a suspicious event. To improve filtering of legitimate events,
we check if at the end of the $T_{s2}$ period, the new link has 
appeared in the opposite direction in the BGP updates received from the monitors and/or local routers.
In other words, if the new link really exists,
then it is probable that it is used also in the opposite direction
and a route (containing the opposite direction) will propagate to a monitor or a local router after some time. The waiting interval $T_{s2}$ can be configured by the operator (speed/accuracy trade-off); here, we select $T_{s2}=5$ minutes, which is enough time for the best BGP paths to converge on most of the monitors~\cite{huston2016bgp}.

\begin{comment}
%%%%%%%%%%   ALBERTO's VERSION   %%%%%%%%%%
Stage 2 introduces a 5-minute delay in exchange for refined filtering
and the ability to estimate the impact of an event marked as
suspicious.  
%In the optional Stage 2, \artemis collects information for an
%additional 5 minutes since the detection of the event under exam.
%
To improve filtering of legitimate events,
we check if at the end of the 5 minute interval, the new link has 
appeared in the opposite direction in the routing
table of any monitor 
or of any local router.
In other words, if the new link really exists,
then it might be possible that it is used also in the opposite direction
and that after BGP convergence a monitor or a local router
will observe such
behavior. We assume the best BGP path to converge, for the given
prefix, within 5 minutes on most of the monitors \cite{huston2016bgp}.
%%%%%%%%%%   end of ALBERTO's VERSION   %%%%%%%%%%
\end{comment}

\myitem{\textit{Stage 2} allows \artemis to further reduce alerts for 
Type-N, $N\geq2$, events.}
The black curve (square markers) in Fig.~\ref{fig:cdf-new-links-times} shows the CDF of the number of events detected as
suspicious at the end of \textit{Stage 2} when using \blue{the public monitors (RouteViews
and RIPE RIS), but not local routers. The improvement only from public monitors is around $1\%$.} 

\blue{However, considering also the local monitors and the impact of the events, significantly increases the gains from Stage 2, as we discuss in the remainder.}
%The improvement is more visible in the lower part of the CDF, where, \eg the percentage of origin ASes that do not see more than 1 alert every 10 days goes from 42.6\% to 43.6\%.
%%%%\textit{Note}: we stress here that the results of Fig.~\ref{fig:rules} are derived through emulation by considering only the \textit{AS-links list from monitors} but not the information from local routers and therefore represent an upper bound for the alert rate.

\myitem{Local routers see significantly more links in the opposite
direction than monitors, thus further improving the filtering of
\textit{Stage 2}.} Using in \textit{Stage 2} the \textit{AS-links list
from local BGP routers} as well, would further reduce suspicious
events. We investigate this effect through simulation: we introduce a
new link in the topology, and after BGP convergence we check whether the new link is seen in the opposite
direction by the local routers. Our results show that the
\textit{local BGP routers} see the opposite
direction of the new link in around $25\%$ (2nd-hop) and $30\%$
(3rd-hop) of the cases, \ie thus filtering 1-2 orders of magnitude
more Type-2 and Type-3 suspicious events compared to the case of using
only the \textit{AS-links list from monitors}% (see Table.~\ref{table:reduction-false-pos-new-links}); in our simulation experiments, both monitors and local routers rarely observe new links in larger distances ($N\geq 4$)
. This rich information that exists locally, highlights further the gains from the self-operated approach
of \artemis.

%\begin{table}[!h]
%\begin{minipage}{\linewidth}
%\centering
%\caption{Simulation results of the reduction (\%) of false positives
%by \textit{Stage 2}, due to the information from monitors and local routers.}
%\label{table:reduction-false-pos-new-links}
%\begin{tabular}{|c|cc|}
%\hline
%{position of new link:}	&{$2^{nd}$ hop}		&{$3^{rd}$ hop}\\
%\hline
%{only monitors}				&{~0.2\%}		&{~4.6\%}\\
%{monitors+local router(s)}	&{24.2\%}		&{31.8\%}\\
%\hline
%%{stream}		&{only monitors}				&{~0\%}			&{~0\%}\\
%%{services}	&{monitors+local router(s)}	&{22.4\%}		&{15.4\%}\\
%%\hline
%\end{tabular}
%\end{minipage}
%\end{table}

\myitem{\textit{Stage 2} provides an estimate of the
impact of the suspicious event.}
Waiting for BGP convergence allows Stage 2 to further 
discover how many monitors see the Type-N suspicious event (\ie the new suspicious link in a route towards the operator's prefix)
and, therefore, estimate the extent of
the ``pollution'' in case the event is a hijack.
When Stage 2 is enabled, \artemis uses this information to trigger
different alert modes and mitigation strategies based on the
configuration provided by the operator (\S~\ref{sec:mitigation-techniques}).

\myitem{\textit{Stage 2} --optionally-- allows the operator to
almost eliminate false positives at the expense of a few false
negatives of negligible control-plane impact.}
The impact (``pollution'') estimate of \textit{Stage 2} can also be
used to further reduce false positives, by raising an alert only if
the number of monitors seeing the event is above a (user-selected) threshold. 
In this way, \artemis can completely ignore a
large number of uninteresting events (\eg legitimate 
changes in routing policies that appear as new links) at the expense of potentially
introducing false negatives that have negligible visible impact on the control plane. 
This is demonstrated in Fig.~\ref{fig:cdf-new-links-times}, which shows that
the majority of the suspicious events we observe in the 
Internet (same experiment as in Fig.~\ref{fig:rules}) are seen by only
\textit{a single} monitor.
%\ed{Probably we should cut this, but I'd like to know what you think
%about this phenomenon: "Thus, often, they are probably the consequence
%of a change in routing policies in the proximity of the monitor AS but
%do not propagate to the rest of the Internet." --ad}

Specifically, according to our experiment in Fig.~\ref{fig:cdf-new-links-times} \blue{(see x-axis for $x\rightarrow 0$)}, by ignoring all new links observed
at only one monitor, \textit{Stage 2} would have generated \blue{\textit{at most one} (or, \textit{zero})} alert in the whole month of March 2017 for \blue{$83\%$ ($63\%$)}
of the origin ASes (green curve). Increasing the threshold further decreases alerts:
if the operator decides to ignore events seen by less than $4$ monitors (blue curve) then the percentage of origin
ASes without at most one (zero) alerts reaches \blue{$94\%$ ($81\%$)}, and for a threshold of $20$ monitors
(red curve) it is \blue{$97\%$ ($90\%$)}. 
%Specifically, according to our experiment in
%Fig.~\ref{fig:cdf-new-links-times}, by ignoring all new links observed
%at only one monitor, \textit{Stage 2} would have \textit{never}
%generated an alert in the whole month of March 2017 for 63\%
%of the origin ASes (green curve). Increasing the threshold further decreases alerts:
%if the operator decides to ignore events seen by less than $4$ monitors (blue curve) then the percentage of origin
%ASes without alerts reaches 81\%, and for a threshold of $20$ monitors
%(red curve) it is $90\%$. 
Finally,
Fig.~\ref{fig:hijack-type-detectability-all} provides an indication of
the rate of potential false negatives this threshold would yield: \eg
for Type-2 hijacks and a threshold of at least $2$ monitors, the
percentage of false negatives (\ie percentage of hijacks with
negligible visible impact on the control plane, seen by exactly one monitor) would be less
than $4\%$.

\section{Mitigation Methodology}
\label{sec:mitigation-techniques}
\blue{Ultimately, a network operator needs to quickly mitigate a hijacking event. To this end, a timely detection is not the only \textit{necessary} condition.} Low false positives, information about the event (\eg estimated impact, relevance of the affected prefix), and an automated system are also key requirements. In this section, we present the \artemis unified approach for detection and mitigation, which satisfies all these conditions, and enables a configurable and timely mitigation.
%The ultimate goal of a network is to quickly mitigate a hijacking event affecting its prefixes. Timely detection of hijacking events is not the only \textit{necessary} condition for their timely mitigation. Low false positives, information about the event (\eg estimated impact, relevance of the affected prefix), and an automated system are also key requirements. In this section, we illustrate the \artemis unified approach for detection and mitigation, which satisfies all these necessary conditions, thus enabling configurable and timely mitigation.

\subsection{ARTEMIS Mitigation Approach}
\textbf{\artemis provides an informative detection of hijacking events
that enables automated and fast mitigation.} The \artemis detection
module can provide the following information --as output-- for each detected event:
\setlist{nolistsep}
\begin{enumerate}
\item affected prefix(es);
\item type of the hijacking event;
\item observed impact (\eg number of polluted monitors);
\item ASN(s) of the AS(es) involved in the event;
\item confidence level (reliability) of the detection.
\end{enumerate}
Note that a detection is always accurate (no false positives; confidence level =  ``certainty'') for 
any type of sub-prefix hijacking events (\cf
\S~\ref{sec:detection-types-other}) and for
exact-prefix Type-0 and Type-1 hijacking events (\cf
\S~\ref{sec:detection-types-0-1}), \ie the events with the highest impact on the control plane. In contrast, the confidence level of an exact-prefix Type-N, $N\geq2$, hijacking event
can be quantified by the result of the detection \textit{Stages 1/2} (\S~\ref{sec:detection-types-N}) 
%the rules $R1,R2$ of Section~\ref{sec:detection-types-N} (\eg $CL(R1_{fail})<CL(R1_{fail}\&R2_{fail})$) 
and allows \artemis to classify an
event as more or less suspicious (\eg confidence level = ``alert by Stage 1'' and/or ``alert by Stage 2'').
%@note i introduced the word "suspicious" on purpose to create a link
%with below --ad

%(note that $R2$ is associated with an operator-defined waiting time threshold).

This rich information is sufficient in most cases for an operator to
decide how to configure the network's reaction to a hijacking or suspicious event. As a
result,
\artemis enables the automation of mitigation: (i) the operator
pre-configures \artemis (mitigation module) to proceed to different
mitigation actions based on the detection output%, via the following mapping:
\blue{; for instance, the following mapping could be used}\footnote{\blue{In this example, the hijack type and hijacker's ASN are wildcards. In a more specific mapping, all five fields of the information presented above could be distinctly used.}}:
\[\text{\{Prefix, Impact, Confidence level\}}\rightarrow\text{Mitigation action};\]
(ii) \artemis executes the pre-selected action \textit{immediately}
after the detection of an event, not requiring manual actions. 

Examples of applying this approach are: (a) the operator selects to
handle an event of limited impact (squatting, few polluted monitors,
etc.) manually instead of triggering an automated mitigation process;
(b) for sensitive prefixes (\eg web-banking), the operator selects
to always proceed to mitigation (\eg even for low-confidence alerts
for Type-N$\geq2$ hijacks), since the cost of potential downtime (or
even compromise in the case of traffic interception attacks) is
much higher than the mitigation cost for a false alert.
%for prefixes associated with sensitive hosts (e.g., web-banking), even if the confidence level of the hijacking alert is not high (e.g., for Type-N, $N\geq2$, events), the operator selects to always proceed to mitigation, since the cost of downtime is much higher than the mitigation cost for a false alert.

\begin{figure}
\centering
\subfigure[]{\includegraphics[width = 0.49\linewidth]{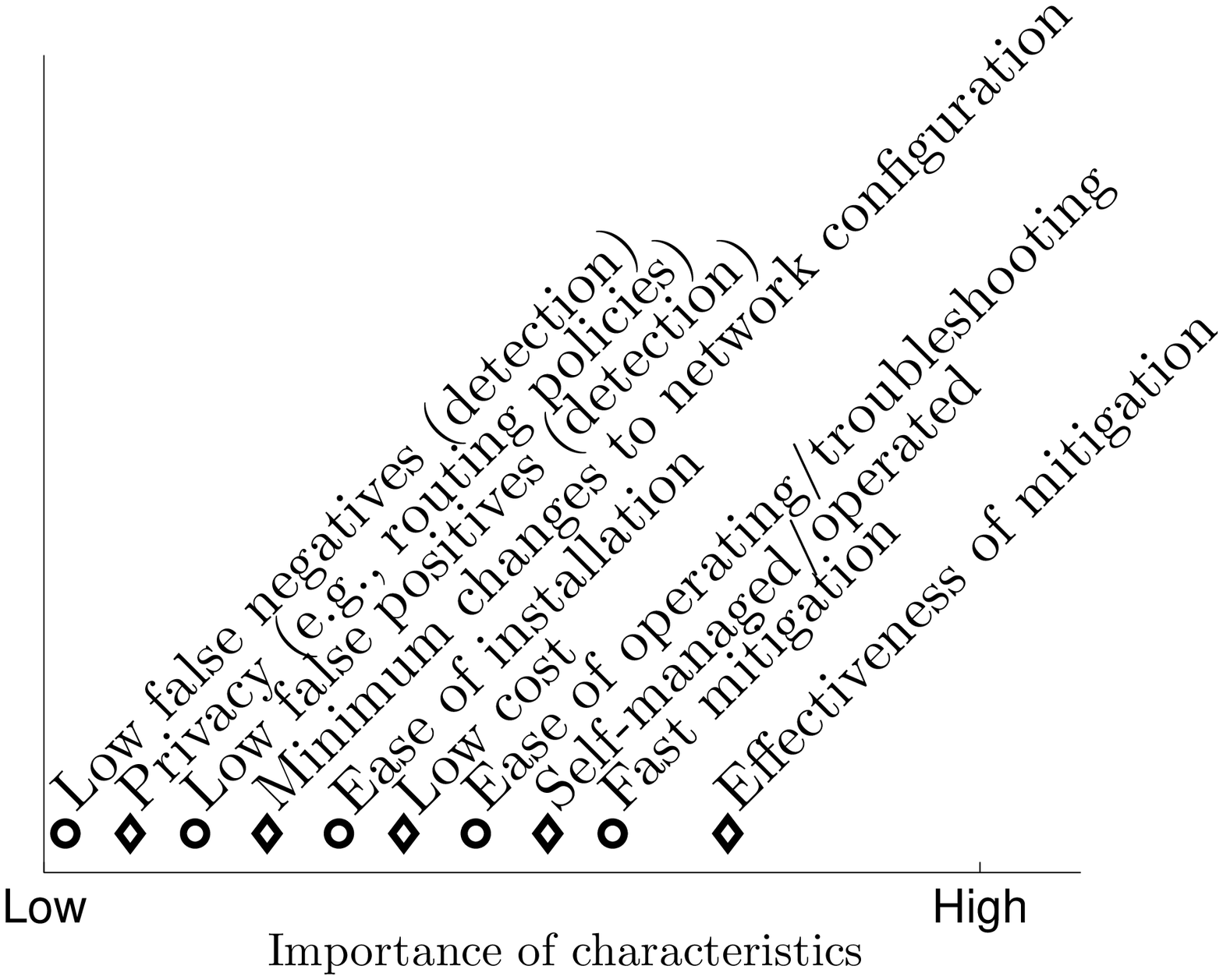}\label{fig:survey-importance-characteristics}}
%%OR
%\subfigure[]{\includegraphics[width = 0.49\linewidth]{./figures/fig_survey_important_factors_mitigation_large_bw1}\label{fig:survey-importance-characteristics}}
%% OR
%\subfigure[]{\includegraphics[width = 0.49\linewidth]{./figures/fig_survey_important_factors_mitigation_large}\label{fig:survey-importance-characteristics}}
\subfigure[]{\includegraphics[width = 0.49\linewidth]{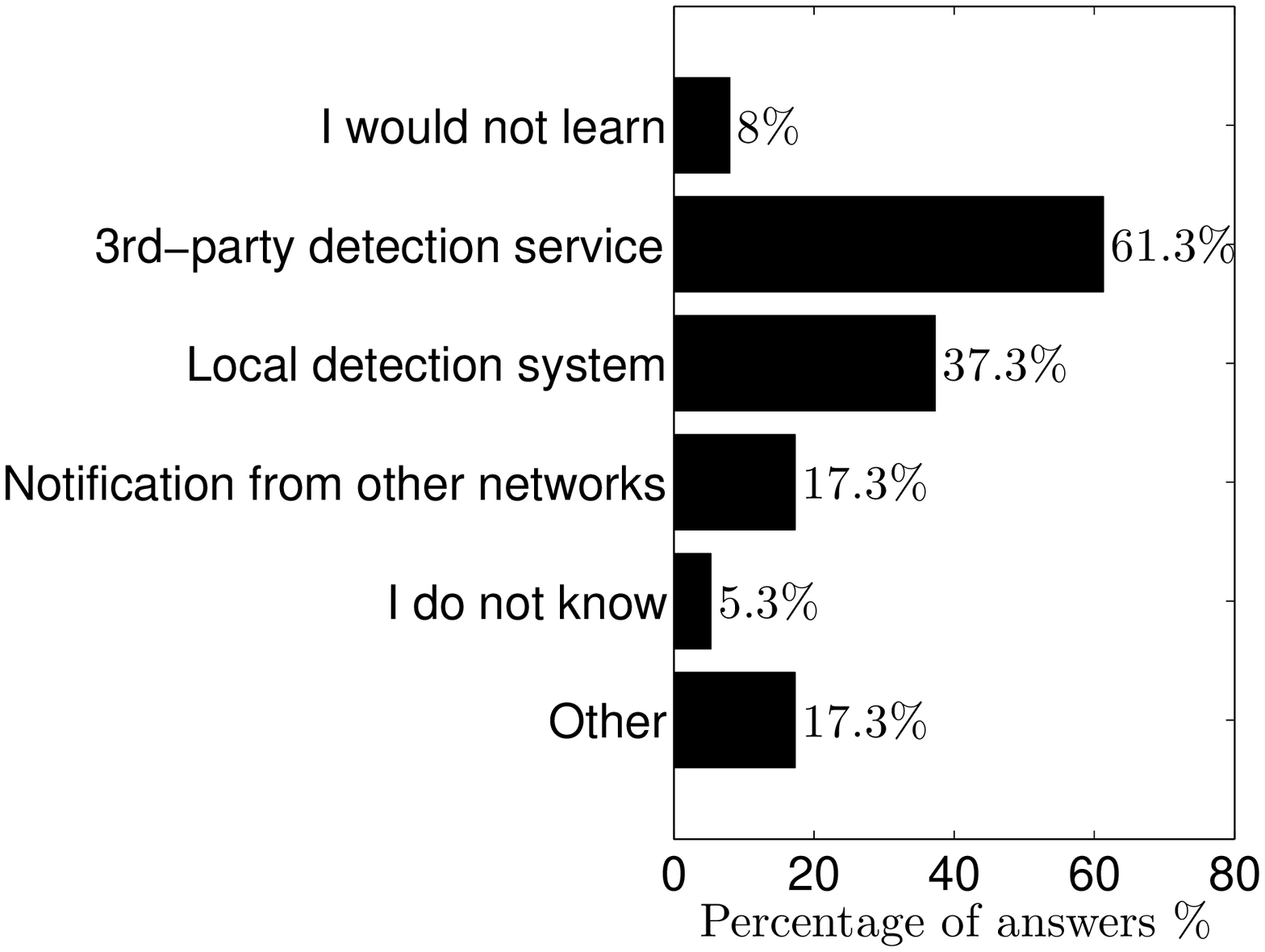}\label{fig:survey-notification}}
\caption{Survey results: (a) ranking of characteristics of a hijacking defense system, based on their importance, by network operators; (b) practices for detecting/learning about hijacking incidents against owned prefixes.}
\end{figure}

\myitem{\textbf{\artemis satisfies operators' needs and outperforms current practices.}} %Our survey 
\blue{We conducted a survey among $75$ network operators (see details in \S~\ref{sec:motivation}) that} shows that the majority of networks rely on third parties for detecting hijacks against their own prefixes (Fig.~\ref{fig:survey-notification}): $61.3\%$ outsource detection to services such as %BGPmon
~\cite{commercial-bgpmon}, and $17.3\%$ expect to be notified by colleagues or mailing lists. 
However, the employment of  third parties may lead to false alerts,
delayed (inferred) detection and thus delayed mitigation (\S~\ref{sec:motivation}).
%However, third parties are not aware of the ground truth, and thus frequently raise false alerts and/or need considerable time to infer a hijacking event (while a single observation is enough for \artemis in most cases), which can significantly delay mitigation. 
In contrast, \artemis provides a reliable and fast detection that also enables fast mitigation, which is one of the main concerns of operators (cf. ``Fast mitigation'' - Fig.~\ref{fig:survey-importance-characteristics}). Moreover, self-operated approaches like \artemis are highly desirable (cf. ``Self-managed/operated'' - Fig.~\ref{fig:survey-importance-characteristics}); we believe that its characteristics (lightweight, no cost for public monitoring services, flexible and configurable) render it ideal for -- at least -- two thirds of the networks not currently employing any local detection system (Fig.~\ref{fig:survey-notification}).
%vassilis: cleaned the redundancy I think
%\ed{i feel there is a bit of redundancy here. i think i cleaned 
%redundancy with intro of 7 by changing that intro (as a
%side effect.. since my purpose was to improve the intro). but still
%things here seem to overlap with intro of 6 (which I wouldn't change)
%and probably stuff in sec. 1 and 2.}

In the following section, we focus on the crucial aspect of the mitigation effectiveness (Fig.~\ref{fig:survey-importance-characteristics}). We study and propose mitigation techniques that build on current practices and can be incorporated in the \artemis approach.

\subsection{Mitigation Techniques}
\blue{We propose two mitigation techniques that can be used with \artemis (other techniques could work as well). Specifically, the victim AS can counteract a hijack with its own resources by \textit{deaggregating} the hijacked prefix (Section 6.2.1), or \textit{outsource} the mitigation to a third party organization, which will announce the prefix on behalf of the victim to reduce the impact of the hijack (Section 6.2.2).}
\subsubsection{Self-operated mitigation with prefix deaggregation}
%\myitem{Self-operated mitigation with prefix deaggregation}.

After receiving an alert for an ongoing hijacking event, operators replied in our survey that they would react by \textit{contacting other networks} ($88\%$ \blue{of the participants}) and/or \textit{deaggregating the affected prefix} ($68\%$ \blue{of the participants}). While the former action involves a significant delay (up to many hours, or even days~\cite{nanog-detection-times}), the latter can be automated and applied immediately after the detection step using the \artemis approach.

Prefix deaggregation
 is the announcement of the more specific prefixes of a certain
prefix. For example, upon the detection of a hijack for the prefix
\textit{10.0.0.0/23}, the network can perform prefix deaggregation and
announce two more-specific sub-prefixes: \textit{10.0.0.0/24} and \textit{10.0.1.0/24}. 
These sub-prefixes will
disseminate in the Internet and the polluted ASes will re-establish
legitimate routes, since more-specific prefixes are preferred by BGP.
Prefix deaggregation is \textit{effective} for \textit{/23} or less-specific
 (\textit{/22, /21, ...}) hijacked prefixes
 (since \textit{/25} prefixes and more-specifics are filtered by most
routers~\cite{Bush-Internet-optometry-IMC-2009}). Moreover, it
can be operated by the network itself without any added cost.
The automation of prefix deaggregation over \artemis is simple, \eg using ExaBGP~\cite{exabgp} or custom scripts \blue{that are triggered immediately after the detection}. A potential mapping could be:
\[\text{\{Prefix length $\blue{<}~/24$, *, *\}}\rightarrow\text{Deaggregation}\]
where * denote wildcards (\ie any impact/confidence level).

To mitigate hijacking events involving \textit{/24} prefixes, in the
following we examine alternative mechanisms, which require the
involvement of additional networks besides the one operating \artemis.

\subsubsection{Outsourcing mitigation with MOAS announcements}
%\myitem{Outsourcing mitigation with MOAS announcements}.

%Community experience 
%and our survey show that networks are reluctant
%to deploy distributed BGP security mechanisms (e.g., BGPsec or RPKI)
%due to the incurred costs, small security benefits (resulting from
%limited adoption), etc. On the contrary, 
%@@above is confusing when mixed with the outsourcing of dos. plus
%it's a repetition of what we said many times. reader will get very
%frustrated
It is common practice for
networks to outsource various security services to a single (or a few)
third-party organization(s). A prominent example is the DDoS mitigation
service offered by organizations to networks that are unable to handle
large volumetric DDoS attacks~\cite{arbor-security-report}. Moreover,
$39\%$ of the participants in our survey do not reject the possibility
to outsource hijacking mitigation.
We also expect that the higher level of accuracy offered by \artemis
and the per-prefix configurability would make more operators consider
outsourcing mitigation, when triggering it is under their control (\eg
allowing them to carefully manage the cost {\em vs} security risk
trade-off).%To this end, 
~We thus propose a mitigation technique that presents several analogies with
the current practice of DDoS mitigation services, and study its
efficiency.

\begin{figure}
\centering
\subfigure[Hijack type 0]{\includegraphics[width=0.49\linewidth]{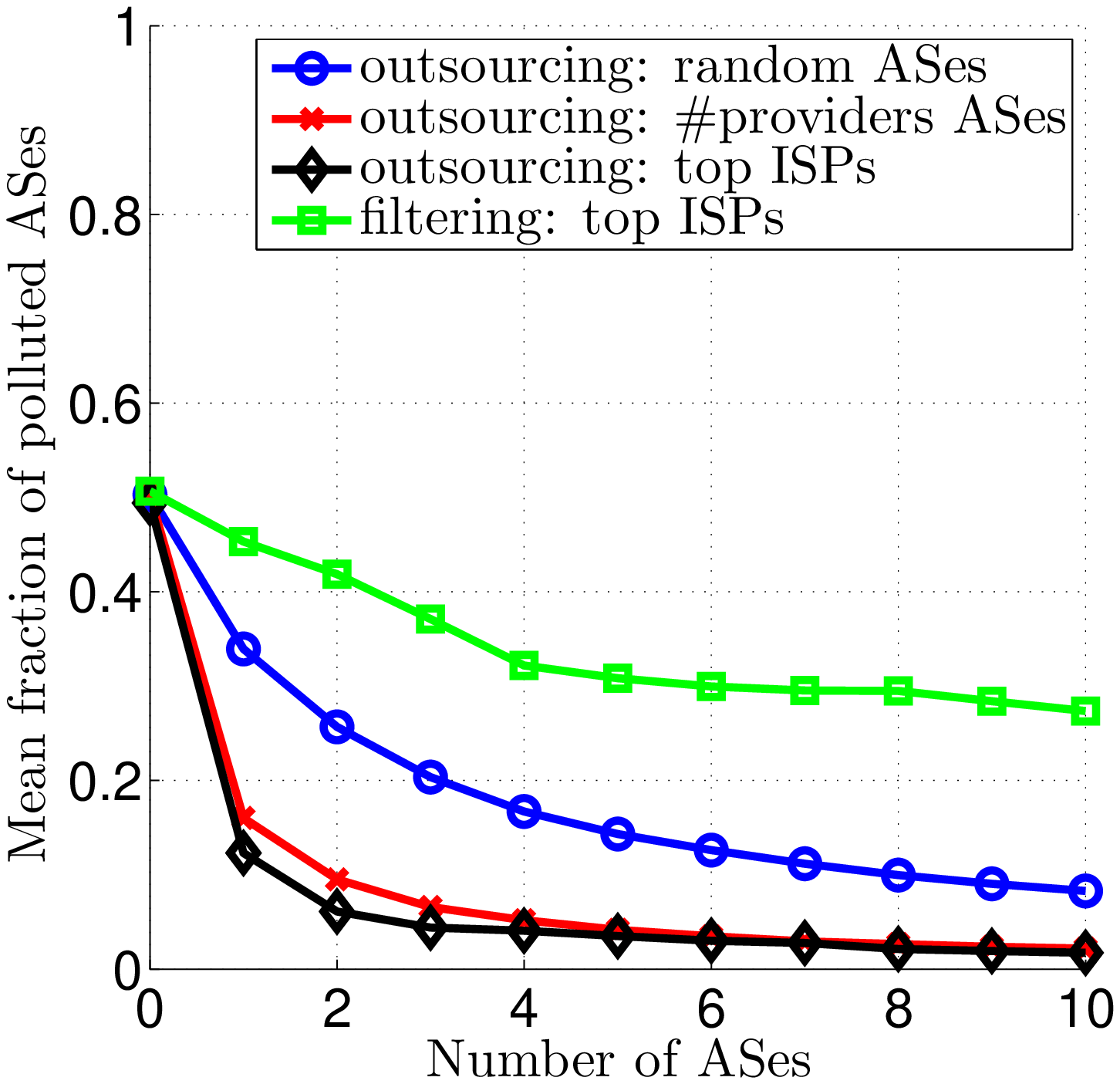}\label{fig:anycast-type-0}}
\subfigure[Hijack type 1]{\includegraphics[width=0.49\linewidth]{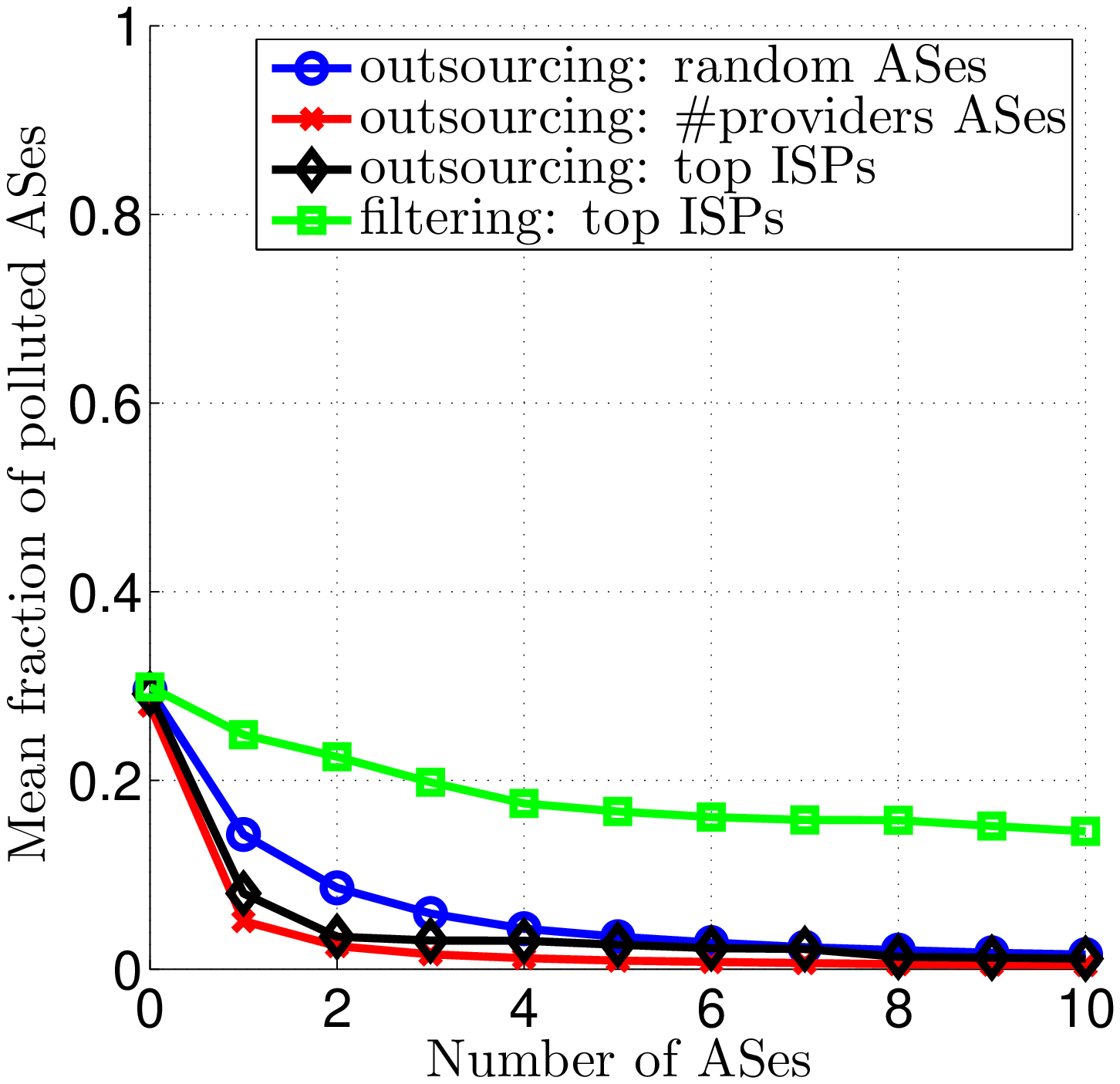}\label{fig:anycast-type-1}}
\caption{Efficiency of mitigation via \textit{outsourcing BGP announcements} to organizations selected (i) randomly, and based on their (ii) number of providers and (iii) customer cone (top ISPs), and via (iv) \textit{filtering} at top ISPs. 
Outsourcing mitigation -even to a single organization- is very effective and significantly outperforms current practices (filtering).
}
\label{fig:anycast}
\end{figure}

%\myitem{\textbf{Outsourcing BGP announcements}}
Outsourcing BGP announcements
is similar to the
outsourced DDoS protection security model, where the organizations
that mitigate the attacks redirect the traffic (using BGP/MOAS or
DNS) to their locations and scrubbing centers, remove malicious
traffic, and forward/relay the legitimate traffic to the victim. In
the case of BGP hijacking, the mitigation organization receives a
notification from the network operating \artemis, and immediately
announces from their location/routers the hijacked prefix. In this
way, the organization attracts traffic from parts of the Internet and
then tunnels it back to the legitimate AS \blue{ (through, \eg MPLS tunnels, direct peering links, or its upstream providers)}. The automation of this
process could be implemented, \eg with \artemis{-triggered} MOAS 
% vassilis: my bad, meant MOAS
%multi-hop BGP sessions \ed{I
%thought we were going to do MOAS. Multi-hop introduces an additional
%hop. Btw, we should cite the paper that evaluated multi-hop when/if we
%mention multihop, without waiting for the relwork section} 
~on the control plane and traffic tunneling on the data plane; a corresponding mapping
in \artemis (potentially only for the most security-sensitive prefixes
owned by the organization) could thus be:
\[\text{\{Prefix length $= /24$, *, *\}}\rightarrow\text{Outsource BGP announcements}\]

More than one external organization can be employed for more effective
mitigation. In the following, we investigate the efficiency of this
technique for different \textit{selection criteria} and
\textit{number} of mitigation organizations. In Fig.~\ref{fig:anycast}
we present simulation results for the remaining number of polluted
ASes (y-axis) after announcing the prefix from different numbers of
mitigation organizations (x-axis) in addition to the network operating
\artemis. We consider three cases where we select the outsourcing organizations (i) \textit{randomly}, and based on their (ii) \textit{number of providers} (which correlates with their mitigation efficiency~\cite{Lad2-Understanding-resiliency-hijacks-DSN-2007}) and (iii) \textit{customer cone} (``top ISPs'') that corresponds to large ISPs~\cite{as-rank-website}. 
%\red{(cf. CAIDA transit degree)}. 

\myitem{\textbf{Outsourcing mitigation even to a single organization
is very effective, and significantly reduces the impact of
hijacking.}} Fig.~\ref{fig:anycast-type-0} shows that outsourcing BGP
announcements to the top ISPs outperforms a selection of ASes with
many providers, while randomly selecting organizations is always less
efficient. However, even a single randomly selected organization can
considerably reduce the impact of the hijacking event (on average),
from $50\%$ to $34\%$ and from $28\%$ to $14\%$ for Type-0
(Fig.~\ref{fig:anycast-type-0}) and Type-1
(Fig.~\ref{fig:anycast-type-1}) events, respectively, which clearly
indicates an effective and robust mitigation technique. Outsourcing to
more than one organization simultaneously and/or carefully selecting
the mitigation organization can further increase the mitigation
benefits, \eg leading to less than $5\%$ polluted ASes  (one order of magnitude lower compared to the initial impact) with only 3 top ISPs for Type-0 events.
%@ we didn't write anything about how practical all this would be in
%reality. but i wouldn't add it now before the deadline

\myitem{\textbf{Outsourcing BGP announcements outperforms current
practices.}} In Fig.~\ref{fig:anycast} we compare the efficiency of
outsourcing against \textit{prefix filtering}, a proactive defense
that needs cooperation of networks and is currently partially deployed
(\S~\ref{sec:motivation}). We consider filtering of the
illegitimate routes from the top ISPs; while filtering applies to origin-AS hijacks today, in Fig.~\ref{fig:anycast-type-1} we assume a potential filtering for Type-1 hijacks as well%\footnote{Note that today filtering protects only against origin-AS hijacks. However, in Fig.~\ref{fig:anycast-type-1} we assume a potential filtering for Type-1 hijacks as well.}
. Our results show that filtering is much less efficient than outsourcing BGP announcements: even with $10$ filtering ASes, the mitigation efficiency is almost equal to (Fig.~\ref{fig:anycast-type-0}) or not better than (Fig.~\ref{fig:anycast-type-1}) using a single randomly selected outsourcing AS. Increasing the number of filtering ASes to a few dozens, barely helps. 

%\textbf{A positive message for the feasibility and deployment of \artemis -like approaches in the near future: existing industry security models can be effective for outsourcing hijacking mitigation.} 
\myitem{\textbf{Existing industry security models can provide highly effective outsourced mitigation.}}
In Table~\ref{table:percentage-anycast-DDoS}, we present the hijacking
mitigation efficiency of different organizations that currently
provide DDoS protection services. We selected, as examples, $5$ organizations of varying sizes\footnote{Namely: Akamai (AK\blue{; ASNs: 20940, 16625}), CloudFlare (CF\blue{; ASN: 13335}), Verisign (VE\blue{; ASNs: 26415, 30060, 7342, 16838}), Incapsula (IN\blue{; ASN: 19551}), and Neustar (NE\blue{; ASNs: 7786, 12008, 19905})% and Level3 (L3)
.}
%, Prolexic (PR), DOSarrest (DA), F5 Networks (F5) and CenturyLink (CL).} 
~and simulated BGP announcements originating from them for the
hijacked prefix. Mitigation with any of them is efficient,
outperforming even top ISPs. Specifically, mitigation from Akamai is
the most efficient, reducing the percentage of polluted ASes to $2.4\%$ (from $50\%$ originally) on average for Type-0 hijacks. This holds also for the other hijack types, where the average percentage of polluted ASes is reduced to $0.3\%$ or less.

\begin{table}
\begin{minipage}{\linewidth}
\centering
\caption{Mean percentage of polluted ASes, when outsourcing BGP
announcements to organizations providing DDoS protection services; these organizations can provide highly effective outsourced mitigation of BGP hijacking.}
\label{table:percentage-anycast-DDoS}
%\begin{small}
\begin{tabular}{|l|ccccccc|}
\hline
{ }&{\hspace{-0.1cm}without}&{\hspace{-0.2cm}top}&{}&{}&{}&{}&{}\\%&{}\\
{ }&{\hspace{-0.1cm}outsourcing}&{\hspace{-0.2cm}ISPs}&{\hspace{-0.1cm}AK}&{\hspace{-0.1cm}CF}&{\hspace{-0.1cm}VE}&{\hspace{-0.1cm}IN}&{\hspace{-0.1cm}NE}\\%&{L3}\\%&{PR}&{DA}&{F5}&{CL}\\
\hline
{\hspace{-0.1cm}Type0}&{50.0\%}&{\hspace{-0.2cm}12.4\%}&{\hspace{-0.1cm}2.4\%}&{\hspace{-0.1cm}4.8\%}&{\hspace{-0.1cm}5.0\%}&{\hspace{-0.1cm}7.3\%}&{\hspace{-0.1cm}11.0\%}\\%&{14.8\%}\\%&{19.5\%}&{21.2\%}&{25\%}&{27.1\%}\\
{\hspace{-0.1cm}Type1}&{28.6\%}&{\hspace{-0.2cm}8.2\%}&{\hspace{-0.1cm}0.3\%}&{\hspace{-0.1cm}0.8\%}&{\hspace{-0.1cm}0.9\%}&{\hspace{-0.1cm}2.3\%}&{\hspace{-0.1cm}3.3\%}\\%&{5.3\%}\\%&{9.7\%}&{10\%}&{10.7\%}&{11\%}\\
{\hspace{-0.1cm}Type2}&{16.9\%}&{\hspace{-0.2cm}6.2\%}&{\hspace{-0.1cm}0.2\%}&{\hspace{-0.1cm}0.4\%}&{\hspace{-0.1cm}0.4\%}&{\hspace{-0.1cm}1.3\%}&{\hspace{-0.1cm}1.1\%}\\%&{1.8\%}\\%&{4.8\%}&{4.9\%}&{3.9\%}&{3.9\%}\\
{\hspace{-0.1cm}Type3}&{11.6\%}&{\hspace{-0.2cm}4.5\%}&{\hspace{-0.1cm}0.1\%}&{\hspace{-0.1cm}0.4\%}&{\hspace{-0.1cm}0.3\%}&{\hspace{-0.1cm}1.1\%}&{\hspace{-0.1cm}0.5\%}\\%&{1.2\%}\\%&{2.7\%}&{2.5\%}&{1.9\%}&{1.9\%}\\
\hline
\end{tabular}
%\end{small}
\end{minipage}
\end{table}

\section{Real-World Experiments}
\label{sec:experiments}
We setup and conduct \textit{real} BGP prefix hijacking experiments in
the Internet (\S~\ref{section:exp-setup}) using the PEERING
testbed~\cite{Schlinker-PEERING-HotNets-2014,peering-website}. We
implemented a prototype of \artemis, which we use to detect and
mitigate the hijacking events, and study the actual \textit{detection
and mitigation times} observed (\S~\ref{section:exp-results}).
%In our experiments, we conduct \textit{real} BGP prefix hijackings in the Internet. Due to the achievable scale of these experiments, their objective is the measurement of actual detection and mitigation times, not measureable on our simulated setup, serving complementarily to our full-scale simulated findings.
%We use the PEERING testbed~\cite{Schlinker-PEERING-HotNets-2014,peering-website}, which provides the capability to announce routable IP prefixes from real ASNs to the rest of the Internet; both the IP prefixes and the ASNs are owned by PEERING, hence, our experiments have no impact on the connectivity of other non-related ASes. We describe our experimental setup in Section~\ref{section:exp-setup} and the associated measurement results in Section~\ref{section:exp-results}.

\subsection{Experimental Setup} \label{section:exp-setup}

\myitem{\artemis prototype.} The current prototype implementation of \artemis interacts with
the streaming services through the RIPE RIS \texttt{socket.io} API 
and \texttt{telnet} for BGPmon. It receives streams of BGP
updates (formatted in plain text from RIPE RIS and XML format from
BGPmon), and keeps/filters only the BGP updates concerning the
network-owned prefixes. %(contained in the local file).
CAIDA's BGPStream will soon support reading from multiple
streaming data sources simultaneously
\cite{bgpstream-paper,bgpstream-v2-beta}
%bgpstream-ietf}
(including RIPE RIS \texttt{socket.io} and BMP feeds, which RouteViews and
others plan to make available at the same time). We envision replacing
the BGP feed
interface of our \artemis implementation using CAIDA's BGPStream API. 
%\eq{I moved this here, but could also go to another section if you think it'd be better --pavlos}

\myitem{Testbed.} \label{section:peering-testbed}
PEERING~\cite{Schlinker-PEERING-HotNets-2014,peering-website} is a testbed that connects with several real networks around the world, and enables its users to announce routable IP prefixes from real ASNs to the rest of the Internet; the IP prefixes and ASNs are owned by PEERING, hence, announcements do not have any impact on the connectivity of other networks.
%PEERING~\cite{Schlinker-PEERING-HotNets-2014,peering-website} is a testbed that enables researchers to interact with the Internet's inter-domain routing system. It connects with several real networks (\textit{PEERING sites}) around the world, and allows to its users to announce routable IP prefixes from real ASNs to the rest of the Internet; the IP prefixes and ASNs are owned by PEERING, hence, announcements do not have impact on the connectivity of other networks.

In our experiments, we use the connections to three real networks/sites (Table~\ref{table:peering-sites}; data of Jun. 2017) %\footnote{The information for each site is from~\cite{peering-website}, as seen in Jun. 2017.} 
that provide transit connectivity to PEERING, which we select due to their Internet connectivity characteristics. GRN and ISI resemble the connectivity of a typical small ISP in the real Internet, while AMS resembles a large ISP.
%In our experiments, we use the connections of PEERING to three real networks/sites (cf. Table~\ref{table:peering-sites}\footnote{The number and ASN of providers and peers for each site are from~\cite{peering-website}, as seen on Jun. 2017.}). We select these sites due to their Internet connectivity characteristics;  the other PEERING sites are either very similar (in terms of connectivity/performance) to some of the employed sites, or have poor visibility as seen from the monitoring services we use. 
We are granted authorization to announce the prefix \textit{184.164.228.0/23} (as well as its two \textit{/24} sub-prefixes), and use the AS numbers \textit{61574} for the legitimate AS, \textit{61575} for the hijacker AS, and \textit{61576} for the outsourcing AS.

\begin{table}
\begin{minipage}{\linewidth}
\centering
\caption{PEERING sites used in the experiments.}\label{table:peering-sites}% \red{(7 Jun. 2017)}
%\begin{small}
\begin{tabular}{|c|cccc|}
\hline
{ID}				&{Network}			&{Location}			&{ASNs}				&{\#peers}\\
{}				&{}					&{}					&{(transit)}			&{(IPv4)}\\
\hline
{\textbf{AMS}}	&{AMS-IX}		&{Amsterdam, NL}		&{12859, 8283 }			&{74}				\\
{\textbf{GRN}}	&{GRNet}				&{Athens, GR}		&{5408}				&{1}				\\
{\textbf{ISI}}	&{Los Nettos}		&{Los Angeles, US}	&{226}				&{1}				\\
\hline
\end{tabular}
%\end{small}
\end{minipage}
\end{table}

% customer cones of providers: 2637-4, 226-21, 12859-18 , 8283-1 

%\subsubsection{Methodology} \label{section:exp-methodology}
\myitem{Methodology.} \label{section:exp-methodology}
%To emulate BGP prefix hijacking events in the real Internet, 
Using the aforementioned ASNs, we create three virtual ASes in PEERING: (i) the legitimate (or victim) AS, (ii) the hijacker AS, and (ii) the outsourcing AS. For each experiment, we connect each virtual AS to a different site/network of Table~\ref{table:peering-sites}, and proceed as follows.
\begin{itemize}[leftmargin=*]
\item[\textbf{1.}] \textit{Legitimate announcement.} The legitimate (victim) AS announces the /23 IP prefix at time $t_{0}$, using ARTEMIS to monitor this prefix for potential hijacking events.

\item[\textbf{2.}] \textit{Hijacking Event.} The hijacker AS hijacks (\ie announces) the /23 IP prefix at time $t_{h} = t_{0}+20min$.

\item[\textbf{3.}] \textit{Detection.} When a hijacked (illegitimate) route arrives at a monitor, ARTEMIS detects the event at a time $t_{d}$ ($>t_{h}$), and immediately proceeds to its mitigation.%, \ie with a delay not greater than $1-2sec$ (due to processing/connection times)

\item[\textbf{4.}] \textit{Mitigation.} The legitimate AS announces
the /24 sub-prefixes (\textit{deaggregation}), or the outsourcing AS announces the /23 prefix (MOAS announcement) at time $t_{m}$ ($t_{m}\approx~t_{d}$). 
\end{itemize}

\myitem{Scenarios.} \label{section:exp-scenarios}
We conduct experiments in several scenarios of different hijacking and mitigation types, considering all combinations of the following parameters:
%In order to investigate the effect of different hijack and mitigation types, and evaluate the performance of our approach in defending against hijacks, we conduct hijacking experiments in several scenarios, by employing all possible combinations of the following (varying) parameters:

\begin{itemize}[leftmargin=*]

\item \textit{Location} (\ie connection to PEERING sites) of the legitimate, hijacker, and outsourcing ASes.

\item \textit{Hijacking event types}: 0 (origin-AS), 1, and 2.

\item \textit{Mitigation} via deaggregation or MOAS announcements.

\end{itemize}

For brevity, we denote a scenario with three letters $\{V,H,M\}$, indicating the location of the \textit{victim}, \textit{hijacker}, and \textit{mitigator} PEERING sites, respectively. For instance, ``\{G,A,I\}'' denotes the experiment where the victim and hijacker ASes are connected to GRN and AMS sites, respectively, and mitigation is performed through BGP announcements from an outsourcing AS connected to ISI. In deaggregation scenarios, the mitigation is self-operated by the victim AS, thus the first and third letters are the same, \eg ``\{G,A,G\}''. When we consider only the hijacking and not the mitigation phase, we use only the first two letters, \eg ``\{G,A,*\}''.

\myitem{Monitoring the Experiments.} \label{section:exp-monitoring}
%\textbf{Real-time monitoring.}
 In the \artemis prototype we use the BGPmon~\cite{bgpmon} and the RIPE RIS~\cite{ripe-ris-real-time} streaming services for the continuous real-time monitoring of the Internet control plane and the detection of hijacking events. In our experiments, we use the same services to monitor the mitigation process as well.

%\textbf{Post-analysis.} 
The BGPStream framework provides BGP updates from \textit{all} the
monitors of RIPE RIS and RouteViews, currently with a delay of several
minutes (see \S~\ref{sec:data}). Hence, we use BGPStream for a
post-analysis of the experiments: after the experiment we collect the
BGP updates received by the monitors during the experiment and
analyze them.
%received by the monitors, \emph{after the experiment has been
%completed}. 
%@"after the experiment has been completed" means we use the updates
%that came after!
We present these results, %in addition to 
\blue{and compare them with} those from the current
real-time monitors, to demonstrate the performance of
\artemis when more monitors turn real-time.

\subsection{Experimental Results} \label{section:exp-results}

We next analyze the results of our experiments, w.r.t.~the \textit{time} needed by \artemis to detect and mitigate hijacking events in various scenarios.
%In this section, we analyze the results of our experiments and measurements, and study the effect of different parameters in the efficiency of hijack detection (cf. Section~\ref{sec:experiment-results-detection}) and mitigation (cf. Section~\ref{sec:experiment-results-mitigation}), as well as in the monitoring efficiency of our employed control-plane sources.

\subsubsection*{Detection} 
\label{sec:experiment-results-detection}
%\myitem{Detection.} \label{sec:experiment-results-detection}
We consider the \textit{detection delay}, $t_{d}-t_{h}$, \ie the time elapsed between the hijacker's announcement ($t_{h}$) and the detection of the event by \artemis ($t_{d}$).
%We first consider the first phase of the experiments, \ie hijack detection, and quantify it using metrics related to the \textit{detectability} and \textit{detection delay} of the hijacks.

\begin{figure}
\centering
\subfigure[detection delay]{\includegraphics[width=0.49\linewidth]{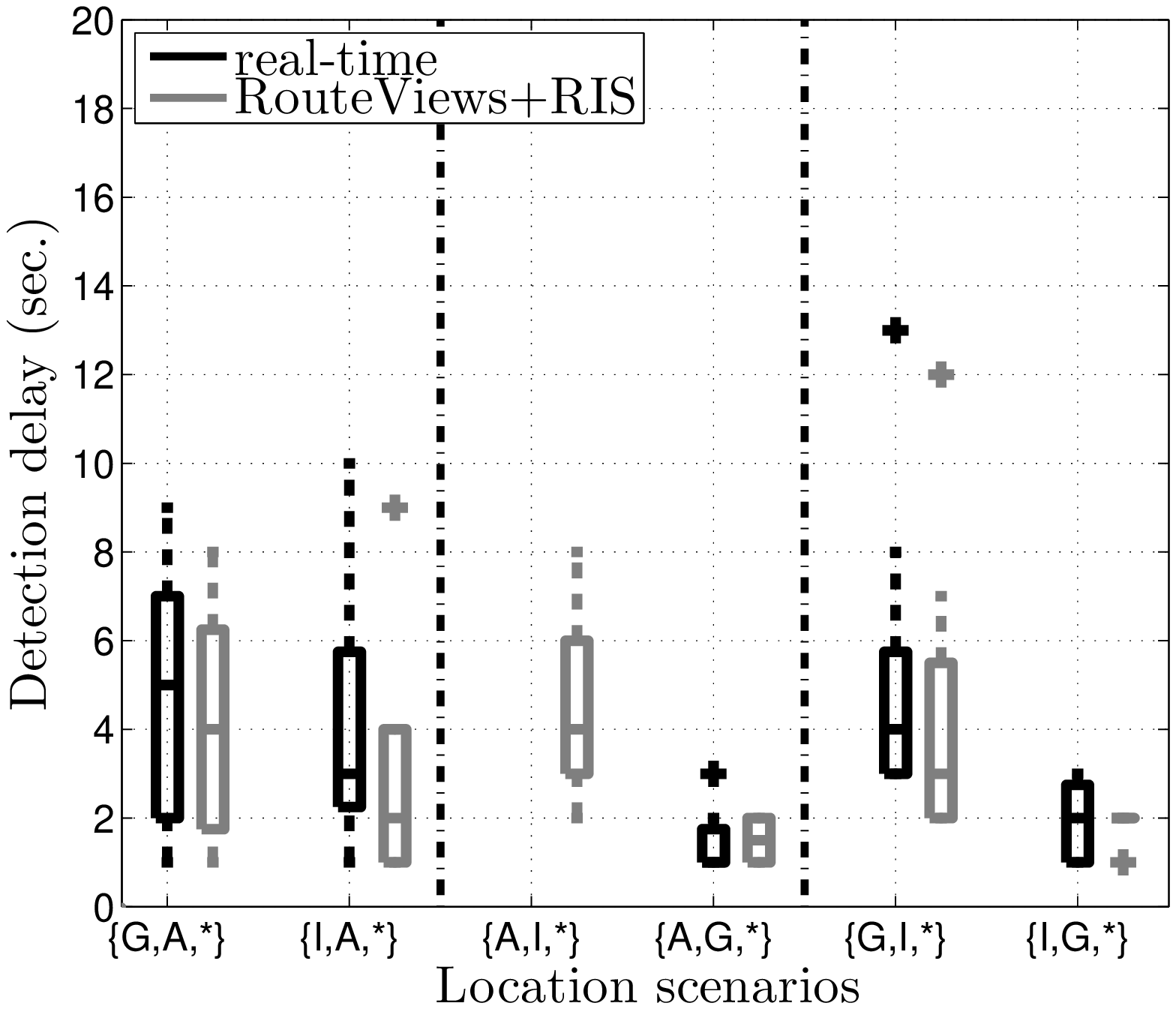}\label{fig:experiments-detection-delay-boxplots-location}}
\subfigure[\# polluted monitors]{\includegraphics[width=0.49\linewidth]{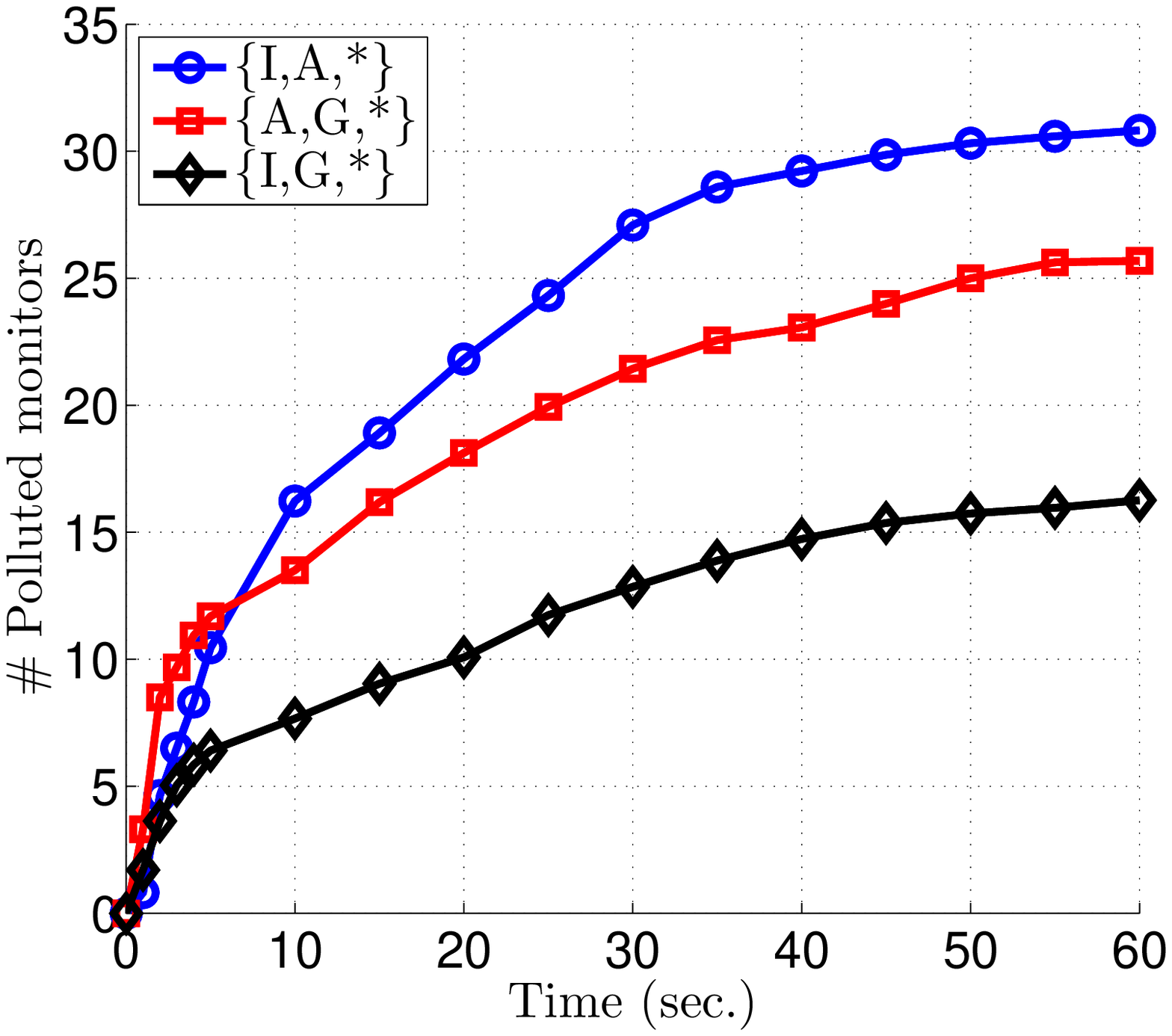}\label{fig:experiments-nb-hijacked-routes-vs-time-only-real-time}}
\caption{(a) Detection delay for different \textit{location scenarios} (x-axis), and origin-AS (type-0) hijacks. (b) Average number of real-time monitors that observed hijacked routes over time.
Boxplots/curves correspond to average values over $10$ experiment runs per scenario.
\artemis detects hijacks within a few seconds (usually $<5$s), while
the hijack is observed by most of the monitors in less than $10$s.}
%Scenarios are presented in groups based on the type of the victim and hijacker ASes (small/large ISPs), as described in the gray labels, separated by the dashed lines.
\label{fig:experiments-detection-delay-boxplots-location-hijacked-routes}
\end{figure}

In Fig.~\ref{fig:experiments-detection-delay-boxplots-location} we present the distribution of the detection delay for different location scenarios, under type-0 hijacking events. Boxplots correspond to the values of $10$ runs per scenario, for either real-time services (black boxplots) or all services (\ie post-analysis with BGPStream for RouteViews and RIPE RIS monitors; gray boxplots). We note that the following insights are valid across hijacking event types, since we observed (results omitted for brevity) that the type does not significantly affect the detection delay; small increases (no more than a few seconds) can though occur because in high-type hijacks, less hijacked routes eventually reach the monitors (due to the preference of shorter AS-paths). Moreover, the tunable waiting time of \textit{Stage 2} (in case \textit{Stage 1} does not suffice, see \S~\ref{sec:detection-types-N}) for type-\{$N\geq2$\} hijacks can be added to the detection delay.

\myitem{\artemis achieves near real-time detection, within a few
seconds of the hijacker's announcement.} \blue{The \artemis detection process is lightweight and thus a hijack event is detected almost instantaneously after the reception of an illegitimate BGP update. Hence, the detection delay is equivalent to the delay of the monitoring services. Specifically, Fig.~\ref{fig:experiments-detection-delay-boxplots-location} shows that the detection} %Detection 
via the
\textit{real-time} services is extremely fast, and in some cases
\textit{only $1$s} is required. In all cases the \textit{median of the detection delay is at most $5$s}. The delay is almost always less than $10$s, and in the worst
case $13$s  (\textit{\{G,I,*\}} scenario in
Fig.~\ref{fig:experiments-detection-delay-boxplots-location}). In
fact, the $1$s delay in some experiments, indicates that the \artemis
approach reduces the detection delay to the propagation time of
BGP updates (from the hijacker to the monitors)\blue{: the detection takes place upon the \textit{first} BGP update that reaches any monitor}. This propagation time
depends on the location/connectivity of the hijacker, \eg we observe that the detection is on average $2-3$s
faster when the hijacker is the GRN site (\textit{\{A,G,*\}} and
\textit{\{I,G,*\}} scenarios).
% We further note that the detection delay only slightly increases (typically less than $10$s) for type-2 hijacks, under $T=0$ for rule R2; when $T>0$, its value is added to the detection delay (we omit the detailed results due to space limitations). 

\myitem{Adding monitors decreases detection delay and increases visibility of hijacks.} If all RouteViews and RIPE RIS monitors provided real-time streams (gray boxplots), detection delay could further decrease; the improvement is small in our experiments, since the detection with real-time services is already fast. Moreover, as already discussed in \S~\ref{sec:understanding-hijacks}, adding more monitors increases the visibility of hijacks. For instance, in the \textit{\{A,I,*\}} scenarios where the victim (AMS) has much higher connectivity than the hijacker (ISI), while the (exact prefix) hijack is not detected by real-time services, using all monitors would enable a timely ($<6$s) detection.

\myitem{Detection is robust.} In
Fig.~\ref{fig:experiments-nb-hijacked-routes-vs-time-only-real-time}
we present the average number of real-time monitors that observed a
hijacked route over time for different scenarios. While \artemis is
able to detect an event from a single (\ie the first seen) hijacked
route, its robustness (\eg against monitor failures) increases with
the number of observed routes. The experimental results in
Fig.~\ref{fig:experiments-nb-hijacked-routes-vs-time-only-real-time}
demonstrate that the detection delay would remain low even under
multiple monitor failures: while the number of observed hijacked
routes differs among scenarios (due to the connectivity of the
hijacker), in all of them \ione more than $5$ monitors observe the
event within $5$s, and \itwo almost half of the monitors that
eventually observe the event, see the hijacked route within $10$s. Our post analysis with BGPstream shows a similar trend (with the respective number of monitors being $3-4$ times higher).

\subsubsection*{Mitigation} \label{sec:experiment-results-mitigation}
We next study how fast the hijacking event is mitigated when using the \artemis approach. To quantify the speed of the mitigation, we define the \textit{recovery delay} as the time elapsed between the pollution of an AS/monitor by a hijacked route, until it receives again a legitimate route (\eg to the deaggregated sub-prefixes).

\begin{figure}
\centering
\subfigure[recovery delay]{\includegraphics[width=0.49\linewidth]{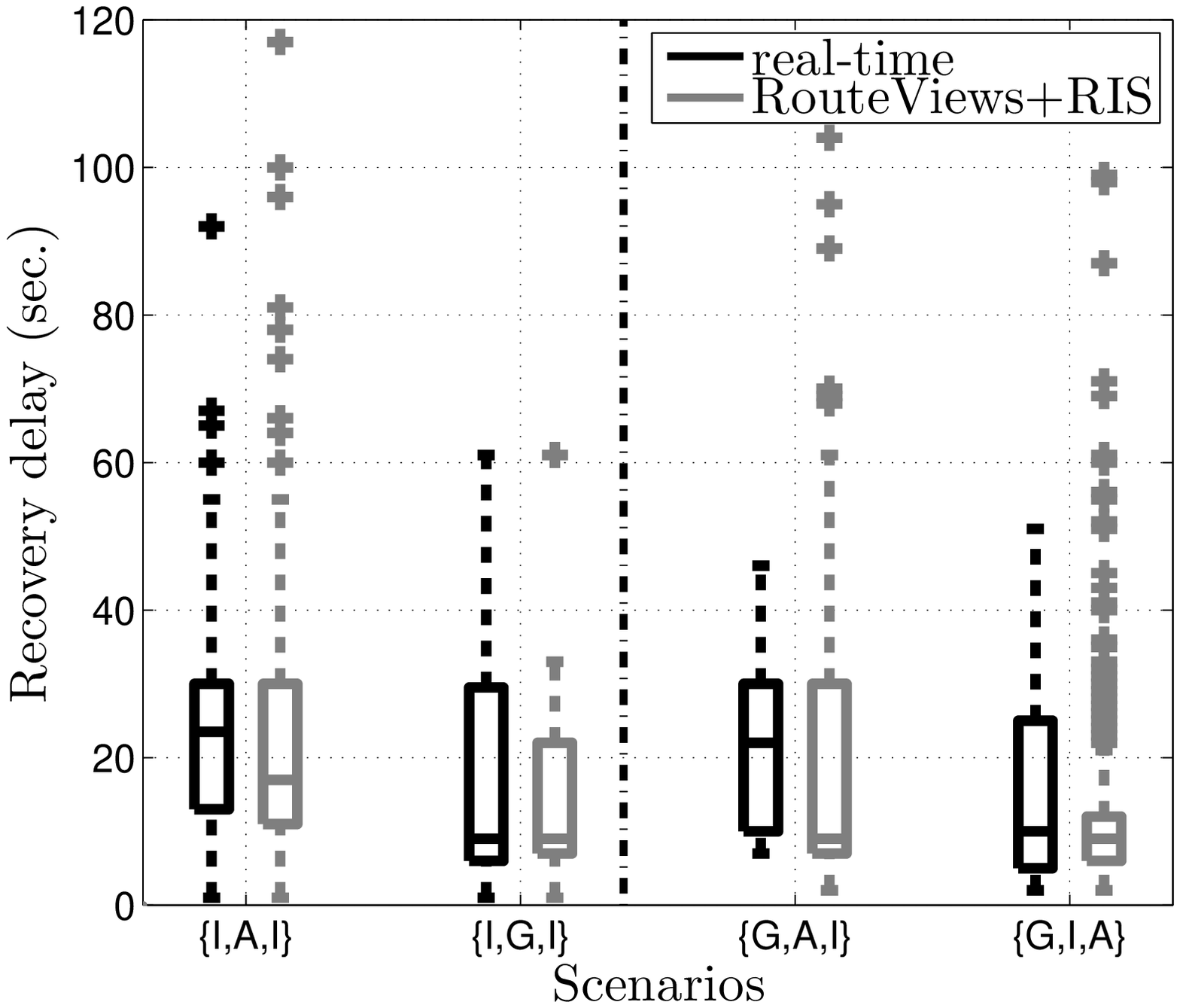}\label{fig:experiments-boxplots-recovery-delay}}
\subfigure[\# polluted monitors]{\includegraphics[width=0.45\linewidth]{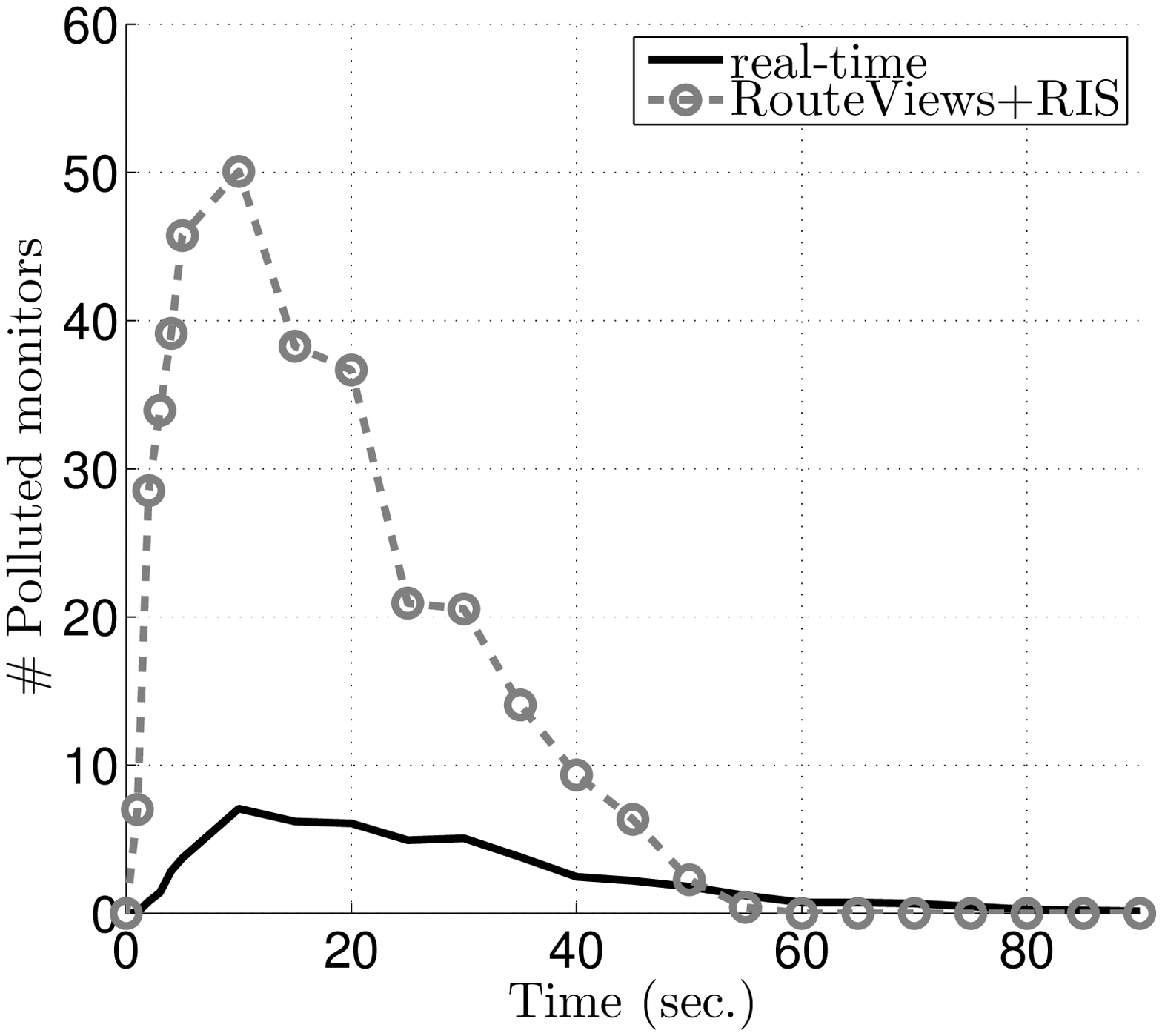}\label{fig:experiments-nb-unrecovered-routes-vs-time}}
\caption{(a) Recovery delay of the the real-time (black) and all (Routeviews+RIS) (gray) monitors; type-0 hijacks and mitigation through deaggregation (\textit{\{I,A,I\}} and \textit{\{I,G,I\}}) and outsourcing BGP announcements (\textit{\{G,A,I\}} and \textit{\{G,I,A\}}) scenarios.  (b) Average number of polluted real-time (black) and all (RouteViews+RIS) (gray) monitors for the \textit{\{I,A,I\}} scenario. Boxplots/curves correspond to average values over $10$ experiment runs per scenario.
With the automated mitigation of \artemis, the vast majority of the ASes recover from the hijack within $60$s.}
\end{figure}

\myitem{\artemis achieves almost complete mitigation of the hijacking event within a minute.} In Fig.~\ref{fig:experiments-boxplots-recovery-delay} we present the distribution of recovery delay (over different ASes and experiment runs) for different location and mitigation scenarios. The time for hijacked ASes to recover is similarly distributed for different mitigation types, and it tends to be higher when the hijacker is a well connected AS (see, \eg \textit{\{I,A,I\}} and \textit{\{G,A,I\}} scenarios where AMS is the hijacker). Nevertheless, the following main observations hold for all scenarios: (i) half of the ASes (see medians) recover from a hijacking event in \textit{less than $30$s}, and the vast majority of them \emph{in less than a minute} (with some outliers reaching up to $2$min.). This clearly demonstrates the benefits of the automated mitigation of \artemis, compared to current practices that usually need several hours to mitigate such events (see \S~\ref{sec:motivation}).

Fig.~\ref{fig:experiments-nb-unrecovered-routes-vs-time} shows the
average number of polluted monitors (\ie with a route to hijacker)
over time for the \textit{\{I,A,I\}} scenario. We observe that the
number of polluted monitors increases fast after the event and reaches
its peak within $10$s. After the event is detected (typically in
$3$-$6$s for the \textit{\{I,A,I\}} scenario; see
Fig.~\ref{fig:experiments-detection-delay-boxplots-location}), the
mitigation starts immediately. The routes for the deaggregated
prefixes start propagating, leading to a fast recovery of the polluted
monitors within $10$-$50$s after the event; \textit{within a minute}
the vast majority of monitors have recovered. We observed similar
behavior in all scenarios, indicating that the real performance of
\artemis in practice would be similar to the results of
Fig.~\ref{fig:experiments-nb-unrecovered-routes-vs-time}.

\section{State of the Art}
\subsection{Real-world Problems with BGP Hijacking}
%\subsection{Motivation and Problem Scope}
\label{sec:motivation}
We now look at the reasons for which BGP prefix hijacking, although
extensively studied, remains a serious threat to Internet operators
and users. To this end, we discuss the current practices and
complement our discussion with findings of a survey we
conducted among network operators.
%\textit{Survey profile:}
The survey~\cite{google-survey-hijacking,report-survey-hijacking} was launched in April 2017, targeting mailing lists of network operator groups to improve our understanding of the: \ione real impact of BGP prefix hijacking, \itwo currently used defenses, as well as \ithree
the concerns, needs and requirements of network operators. We received answers from $75$ participants operating a broad variety of networks (ISPs, CDNs, IXPs, etc.) around the world (the detailed results of the survey can be found in~\cite{report-survey-hijacking}). 

\myitem{\textbf{Operators are reluctant to deploy proactive defenses, since they offer limited protection against hijacking.}}
Several modifications to BGP to protect networks against prefix hijacking have been proposed~\cite{Kent-secure-BGP-JSAC-2000,Subramanian-listen-whisper-NSDI-2004,bgpsec-specification-2015,Karlin-PGBGP-ICNP-2006}, but are not implemented due to several political, technical, and economic challenges. Proactive defenses that are deployed mainly comprise prefix filtering and RPKI~\cite{how-secure-goldberg-ComNet-2014,Goldberg-why-so-long-ACM-Comm-2014,RPKI-deployment-2016}. 
\textit{Prefix filtering} can be used by ISPs to
discard route announcements for prefixes that their customers are not
allowed to originate. However, prefix filtering is currently applied
by a small number of ISPs (due to lack of incentives, poor trust
mechanisms, need for manual maintenance,
etc.~\cite{Goldberg-why-so-long-ACM-Comm-2014}) and offers protection
only against a few potential hijackers (\ie their customers) and hijacking events (origin-AS).
\textit{RPKI} enables automated route origin authentication (ROA) in the Internet, to prevent origin-AS hijacks. However, the small percentage of prefixes covered by ROAs (around 7\% in Oct. 2017\cite{rpki-monitor}) and the limited deployment of RPKI route origin validation (ROV)~\cite{rpki-monitor,RiPKI-hotnets-2015,RPKI-deployment-2016} leaves the vast majority of networks unprotected~\cite{Jumpstarting-BGP-sigcomm-2016,RPKI-deployment-2016}. Our survey results and previous studies~\cite{RPKI-deployment-2016} reveal the main reasons that hinder the deployment of RPKI: little security benefits, mistrust issues, inter-organization dependencies for issuing ROA certificates, operating costs, and complexity.

\myitem{\textbf{Hijacking events, under current practices, have a lasting impact on the Internet's routing system.}}
Due to the insufficiency of proactive mechanisms, networks mainly defend against hijacking events in a \emph{reactive} manner.
% that comprises two steps: \emph{detection} and \emph{mitigation}
The speed of the reactive defenses is crucial; 
%since the Internet is affected until the hijacking is mitigated, 
even short-lived events can have severe consequences~\cite{hijack-BitCoins}. However, the reality shows that, currently, hijacking events are not quickly mitigated. 
For instance, back in 2008, a hijacking event affected YouTube's
prefixes and disrupted its services for $2$
hours~\cite{ripe-pakistan}. More recently, in Sep.~2016, BackConnect
hijacked, at different times, several ASes; the events lasted for
several hours~\cite{backconnect-hijack-2016}.
%@space saving
% In Jan. 2017, the Iranian state telecom TIC hijacked disparate pornographic websites for more than a day~\cite{iranian-hijack-2017}.
In Apr. 2017, financial
services, like Visa and Mastercard, and security companies, like
Symantec, were hijacked by a Russian company for seven
minutes~\cite{russian-hijack-2017}. 
%Besides, it is also important to note that these hijacks always involved /24 prefixes.
Moreover, past experience of operators who participated in our survey shows that their networks were affected for a long time by hijacking events: more than $57\%$ of events lasted \textit{more than an hour}, and $25\%$ lasted even more than a day; only $28\%$ of the events were short-termed, lasting a few minutes ($14\%$) or seconds ($14\%$).
%This is further stressed by findings that show more than $20\%$ of hijacking events last less than $10min.$~\cite{Shi-Argus-IMC-2012}, and that even short-lived events can have severe consequences~\cite{hijack-BitCoins}.

\myitem{\textbf{The mitigation of hijacking events is delayed primarily due to third-party detection and lack of automation.}}
Reactive defenses comprise two steps: \emph{detection} and
\emph{mitigation}. Several systems have been proposed for prefix
hijacking
detection~\cite{Chi-cyclops-CCR-2008,Lad-Phas-Usenix-2006,Zhang-Ispy-CCR-2008,Zheng-lightweight-hijacks-2007,Shi-Argus-IMC-2012,Hu-accurate-hijacks-SP-2007,heap-jsac2016},
with most of them being designed to operate as third-party services;
they monitor the Internet control/data plane and upon the detection of
a suspicious event or anomaly, notify the involved ASes. Our survey
reveals a similar trend in practice: more than $60\%$
rely on third-parties (\eg~\cite{commercial-bgpmon}) to get notified about suspicious events involving their prefixes. Although state-of-the-art third-party detection services can quickly notify networks about suspicious events\footnote{However, $17\%$ of the participants in our survey expect to get notified for hijacking events by receiving notification from colleagues, clients, mailing lists, etc., which implies significantly delayed detection.},
the alerts are not always accurate (\ie false positives), as
discussed in~\cite{nanog-detection-times} and self-reported in our
survey~\cite{report-survey-hijacking}. %, even for popular services. %~\cite{commercial-bgpmon}. 
False alerts might be triggered by third-parties for legitimate events (\eg MOAS, traffic engineering, change of peering policies),
due to missing/inaccurate/stale information. 
%@below i don't think is needed since we already explain in the intro.
%\blues{For instance, when a network announces a sub-prefix for traffic engineering purposes without having earlier notified the third-party service, then a false alert for a sub-prefix hijacking might be raised (false positive). Another real-world example, is that the third-party may decide to not raise alarms for sub-prefixes, since the client is sloppy in updating the list (which will generate false negatives for sub-prefix hijacking attacks).} 
As a result, operators need to manually verify the alerts received by third party services; this process introduces significant delay in the detection step, and prevents networks from automating their mitigation counteractions. Finally, extra delay is added to the process of mitigation itself, which frequently takes place in an ad-hoc way: for example, upon the detection of a hijack, operators start contacting other operators to identify the root of the problem, and resolve it. Interestingly, this is the only action that $25\%$ of operators in our survey would take to mitigate the hijack; however, with this approach the resolution of the problem might require several hours or even days~\cite{nanog-detection-times}.

\subsection{Related Literature}
\label{sec:related}
%\artemis is a detection and mitigation system which 
%operates on the control plane. To the best of our 
%knowledge, there are no other related systems which integrate reliably
%the detection and mitigation cycles. Therefore, we compare
%our work to detection systems (Section~\ref{sec:relwork-detection}) 
%and mitigation mechanisms (Section~\ref{sec:relwork-mitigation})
%separately, to highlight the similarities and differences of \artemis
%with their approaches.

\subsubsection{Detection of BGP Hijacking}
\label{sec:relwork-detection}

BGP hijacking detection approaches can be classified based on the type of information they use, as: (i) control-plane, (ii) data-plane, and (iii) hybrid. Each type has its own strengths and weaknesses, which we analyze in the following. %For convenience, we also summarize our findings regarding the classes of hijacking events detected by the different detection systems (including \artemis) in Table~\ref{table:taxonomy-related-work}.
For convenience, we also summarize in Table~\ref{table:taxonomy-related-work} the classes of hijacking events that can be detected by existing systems.

Similarly to \artemis, \textit{control-plane approaches}~\cite{commercial-bgpmon,Chi-cyclops-CCR-2008,Lad-Phas-Usenix-2006},\blue{\cite{qiu2007detecting}} collect BGP updates or routing tables from a distributed set of BGP monitors and route collectors, and raise alarms when a change in the origin-AS of a prefix, or a suspicious route is observed.
%\footnote{\blue{Note that~\cite{qiu2007detecting} focuses primarily on the detection of policy violations. However, this induces false positives since many such events can be legitimate (e.g., establishment of new peering links, or one-sided change of routing policies). In fact, complex per-prefix policies, beyond e.g., valley-free assumptions, or the invocation of backup legitimate links~\cite{chen2009sidewalk}, may exacerbate this problem. Thus, ARTEMIS concerns itself with entirely new legitimate links vs. fake links, and not policy violations taking place over an existing link, which induce heavy ambiguity.}}. 
Since they passively receive BGP feeds, they are considered quite lightweight. They can detect Type-0 (and Type-1) hijacking events%\footnote{Assuming that the needed ground truth is supplied by the operator using them, with respect to its first-hop neighbors and MOAS practices.}
, both for exact prefixes and subprefixes, independently of how the hijacker handles the attracted traffic on the data plane (blackholing-\textit{BH}, imposture-\textit{IM}, man-in-the-middle-\textit{MM}). However, in contrast to \artemis, state-of-the-art systems~\cite{commercial-bgpmon} miss advanced type-N, N$\geq$2 hijacking events that are harder to detect and can be used
by a sophisticated attacker.
%, in the context of an attack, to guide the traffic through a stealthy Man-in-the-Middle. 
Furthermore, since they are designed as third-party detection services, they have to deal with the real-world problem (\S~\ref{sec:motivation}) of keeping what they observe consistent with the ground truth on the operator's side, to achieve low false-positive rates while preserving their real-time performance.% This is a very complicated objective for third parties.

\textit{Data-plane
approaches}~\cite{Zhang-Ispy-CCR-2008,Zheng-lightweight-hijacks-2007}
follow complementary methods to \artemis, using pings/traceroutes
to detect hijacks on the data plane. They continuously monitor the
connectivity of a prefix and raise an alarm, when significant changes
in the reachability~\cite{Zhang-Ispy-CCR-2008} of a prefix or the
paths leading to it~\cite{Zheng-lightweight-hijacks-2007} are
observed. iSpy~\cite{Zhang-Ispy-CCR-2008} can be deployed by the network operator
itself (similar to \artemis). However, it cannot reliably
detect sub-prefix hijacking events, since it targets few IP
addresses per prefix, and can be severely affected by temporary link failures or congestion near the victim's network, increasing its false positive rates. %Attacks targeting sub-prefixes are not hard to perform and can have persistent large-scale effects, e.g., when combined with \textit{MM} traffic manipulation~\cite{defcon16-attack}. 
Finally, since data-plane approaches require a large number of active measurements to safely characterize an event as a hijack, they are more heavyweight than control-plane-assisted approaches~\cite{Shi-Argus-IMC-2012}. 

\textit{Hybrid approaches}~\cite{heap-jsac2016,Shi-Argus-IMC-2012,Hu-accurate-hijacks-SP-2007},\blue{\cite{qiu2009locating,schlamp2013forensic}} combine control and data plane information, and sometimes query external databases (\eg Internet Routing Registries, IRR)~\cite{heap-jsac2016,Shi-Argus-IMC-2012},
%\footnote{HEAP~\cite{heap-jsac2016} and Argus~\cite{Shi-Argus-IMC-2012} also query external databases, such as Internet Routing Registries (IRR) to retrieve auxiliary information \eg import/export routing policies or organisational information. However, both works advocate the need for careful validation since IRR databases can be outdated or even hold conflicting information.} 
to detect multiple classes of hijacking events. HEAP~\cite{heap-jsac2016} can detect any type of \texttt{AS-PATH} manipulation on the control plane, but is limited to sub-prefix hijacking events which result in blackholing or imposture on the data plane. Thus, it misses \textit{MM} attacks both for exact prefix and sub-prefix hijacks.
The state-of-the-art detection system Argus~\cite{Shi-Argus-IMC-2012}, is able to achieve few false positives/negatives and timely detection, both for exact prefixes and subprefixes, by correlating control and data plane information. However, Argus considers only \textit{BH} %and no \textit{IM} or \textit{MM} hijacking events,
attacks, 
whereas \artemis is able to detect a hijack even if the hijacker relays traffic (\textit{MM}) or responds (\textit{IM}) to it.
The same issue is faced by~\cite{Hu-accurate-hijacks-SP-2007}, where only \textit{BH} and \textit{IM} attacks can be detected for any kind of prefix, while \textit{MM} attacks remain under-the-radar.
\blue{LOCK~\cite{qiu2009locating} locates attacker ASes by actively monitoring control/data-plane paths towards the victim prefix. It relies on the evaluation of AS adjacencies to detect \textit{BH}, \textit{IM} or \textit{MM} attacks, but it might miss sub-prefix and stealthy Type-U hijacks. Schlamp \emph{et al}~\cite{schlamp2013forensic} analyze and focus on a specific hijack case where certain conditions, such as the attack on unannounced BGP prefixes (BGP squatting), apply; data sources such as IRRs or DNS could be used to warn vulnerable ASes.}
%This highlights a general limitation of approaches based on or assisted by data-plane measurements, as well as previous control-plane approaches~\cite{Chi-cyclops-CCR-2008,Lad-Phas-Usenix-2006}; under-the-radar \textit{MM} attacks can have severe consequences~\cite{russian-hijack-2017,defcon16-attack}.

\blue{Finally, while there is no consistent ground truth or dataset
with which to compare all the claimed FN/FP rates of the aforementioned approaches, we stress that the detection approach of \artemis (\S~\ref{sec:detection}, summarized in Table~\ref{table:taxonomy-detection}) is the first to combine \textit{all} the following characteristics:} self-operated, ground truth-based, lightweight detection, allowing for increased accuracy of alerts (0 false positives for most classes, virtually 0 false negatives), and comprehensiveness in terms of attack class coverage, no matter how the attacker manipulates the control and data planes to execute the hijack.

\subsubsection{Mitigation of BGP Hijacking}
\label{sec:relwork-mitigation}

Several proposals exploit cryptographic mechanisms to prevent BGP hijacking~\cite{rpki-rfc,bgpsec-specification-2015,Kent-secure-BGP-JSAC-2000,Subramanian-listen-whisper-NSDI-2004}. Others~\cite{Karlin-PGBGP-ICNP-2006} delay the installation of suspicious BGP routes, in order to allow network administrators to verify first and then install them. However, these approaches require modifications to BGP and/or global adoption, as proactive countermeasures to hijacking events; this has been shown to be infeasible due to important technological, political and economic factors. 
In contrast, we propose reactive self-operated mitigation (prefix deaggregation) or outsourcing it to a single (or, a few) organization(s), which are based on security models used in practice and -as shown in our study- can be very efficient, without requiring large-scale coordination. 
%Compared to these efforts, our approach uses prefix deaggregation and BGP (MOAS) outsourcing as immediately deployable, automatic and efficient measures to mitigate prefix hijacking, without requiring global coordination. 
In fact, we show (see Fig~\ref{fig:anycast-type-0}) that using only a handful top ISPs for outsourcing BGP announcements, the attained benefit ($<5\%$ attacker success rate) would require two orders of magnitude more top ISPs to coordinate and perform Route Origin Validation (ROV) in RPKI~\cite{RPKI-deployment-2016}. 

Zhang \emph{et al.}~\cite{Zhang-practical-defenses-CoNext-2007} propose a reactive mitigation mechanism based on the purging of illegitimate routes and the promotion of valid routes. 
Compared to outsourcing BGP announcements, the approach of~\cite{Zhang-practical-defenses-CoNext-2007} requires one order of magnitude more mitigator ASes (``lifesavers'') to achieve similar benefits, as well as complex coordination among these ASes.
A similar approach to outsourcing BGP announcements is introduced in~\cite{qiu2010towerdefense}, whose focus is on selecting an optimal set of ASes as monitor/relay ``agents'' per victim-hijacker pair. Those results are complementary to our study which considers \emph{existing} monitoring infrastructure and organizations that \textit{currently} offer outsourced security services.

\section{Conclusions}
\label{sec:conclusion}
BGP prefix hijacking, based on accidental misconfiguration or malicious intent,
is a problem that continuously pests Internet organizations and 
users, resulting in high-profile incidents. State-of-the-art solutions, proposed in
research or adopted in daily operations, are not able to counter this situation due
to issues related to: 
\ione attacker evasion (\ie comprehensiveness of detection, e.g., for MitM attacks), 
\itwo problematic accuracy of detection alerts (resulting
in 
\ithree slow manual verification and mitigation processes), and 
\ifour incompatibility with the real-world
needs of network operators for information privacy and flexibility of countermeasures.

%Despite an abundance of different flavors of approaches for detecting
%and mitigating the problem (in separate phases), either proactively or
%reactively, it is still associated with detrimental effects on today's networks.

%For example, state-of-the-art third-party detection mechanisms, usually employed
%by network operators, suffer from multiple issues,
%with the most important being accuracy in terms of false positives/negatives,
%as well as comprehensiveness in terms of advanced hijacking event types,
%such as MitM.
%Sophisticated/complicated approaches that in theory may alleviate some of these
%issues, are not adopted in practice, thus preventing efficient and timely mitigation
%of the problem at Internet scales.
%}

In this work, we proposed \artemis, a \textit{self-operated} control-plane approach
for defending against BGP prefix hijacking. \artemis departs from the
common approach of third-party detection/notification systems, and
exploits local information and real-time BGP feeds from publicly
available monitoring services in order to provide an \textit{accurate}, \textit{comprehensive}
and \textit{timely} detection. This detection approach enables a
\textit{potentially automated}, \textit{configurable}, and \textit{timely}
mitigation of hijacking events, that satisfies the needs and
requirements of operators (as \eg expressed in our survey) and is highly effective, based on currently used practices and outsourcing security models. 
Moreover, as part of our study, we demonstrated the high capability of public monitoring infrastructure for hijacking detection, and showed that the planned transitions to more pervasive real-time streaming can bring substantial benefits. Our simulation results support the feasibility of the \artemis approach while our real-world experiments show that it is possible to neutralize the impact of hijacking attacks within a minute, a radical improvement compared to the defenses used in practice by networks today. 

%In terms of coexistence with current technologies, \artemis can operate in parallel with RPKI (\eg de-aggregated prefixes and ASNs of outsourcing organizations can be added to ROAs), and could even be used in cooperation with RPKI (\eg updating or verifying the RPKI database). Moreover, building on its modular design, it can be enhanced with detection/mitigation features against attackers beyond the adopted threat model (\ie hijackers controlling multiple ASes, launching Type-N, $N\geq2$, attacks).

%Finally, the \artemis approach paves the way for autonomous,
%reliable and fast joint detection and mitigation systems, using rich
%information from control-plane monitoring services. In particular, it can assist in 
%understanding potential attacks against BGP deployments 
%(\eg making more hijacks detectable in real-time, and shedding 
%light on the notorious MitM attacks), thus informing the design 
%and deployment of more secure inter-domain routing protocols 
%and/or new defense systems.

\section*{Acknowledgements}
We would like to thank Henry Birge-Lee for his extremely useful feedback while preparing this paper.

This work was supported by the European Research Council grant agreement no. 338402, the National Science Foundation grant CNS-1423659, and the Department of Homeland Security (DHS) Science and Technology Directorate, Cyber Security Division (DHS S\&T/CSD) via contract number HHSP233201600012C.

\bibliographystyle{plain}
%\bibliographystyle{abbrv}
%\bibliographystyle{ieeetr}
%\bibliographystyle{acm}
%
%\small
%\bibliography{artemis,outages,longnames,net,rfc}
%\bibliography{SDN_hijacks}
% If above is commented out and the next line is not, then we're using
% biblatex+biber and ***WE NEED TO UPDATE THE .bib IN preamble.tex ***

\balance
%\printbibliography

\end{document}